\documentclass[useAMS,usenatbib,usegraphicx]{mn2e}

\usepackage{amsmath}
\usepackage{txfonts}
\usepackage{color}
\usepackage{array}
\usepackage{graphicx}
\usepackage{epstopdf}
\usepackage{color}
\usepackage{bm}
\usepackage{cmap}
\usepackage{cleveref}
\usepackage{booktabs}
\usepackage{tabularx}
\usepackage{natbib}
\usepackage{multicol}
\usepackage{float}
\usepackage{subfloat}

\newcommand{\nup}{\nu_{\mathrm{p}}}
\newcommand{\nua}{\nu_{\mathrm{a}}}

\begin{document}

\title[Electron transport through nuclear pasta]{
Electron transport through nuclear pasta in magnetized neutron stars}

\author[D. G. Yakovlev]{
D. G. Yakovlev\thanks{E-mail: yak.astro@mail.ioffe.ru}
\\
Ioffe Physical Technical Institute, 26 Politekhnicheskaya,
St.~Petersburg 194021, Russia}

\date{Accepted . Received ; in original form}
\pagerange{\pageref{firstpage}--\pageref{lastpage}} \pubyear{2014}
\maketitle \label{firstpage}

\begin{abstract}
We present a simple model for electron transport in a possible layer
of exotic nuclear clusters (in the so called nuclear pasta layer) 
between the crust and liquid core of a strongly 
magnetized neutron star. The electron transport there can be strongly
anisotropic and gyrotropic. The anisotropy is produced by different
electron effective collision frequencies along and across local symmetry axis
in domains of exotic ordered nuclear clusters and by
complicated effects of the magnetic field. We also calculate
averaged kinetic coefficients in case local domains are freely
oriented. Possible applications of the obtained results and open
problems are outlined.
\end{abstract}


\section{Introduction} \label{sec:intro}

Neutron stars are thought to consist of the outer crust, inner crust
and the core (e.g., \citealt{ST83,HPY07}). The outer crust extends
to densities $\rho \leq \rho_{\rm ND}$, where $\rho_{\rm ND} \sim
(4-6) \times 10^{11}$ g~cm$^{-3}$ is the neutron drip density. It
mostly contains strongly degenerate electrons and atomic nuclei
(remnants of atoms fully ionized by the huge electron pressure).
When the density increases, the nuclei become more neutron rich.

The inner crust (where $\rho_{\rm ND} \leq \rho \leq \rho_{\rm cc}$,
$\rho_{\rm cc} \approx \rho_0/2$ being the density at the crust-core
interface and $\rho_0 \approx 2.8 \times 10^{14}$ g~cm$^{-3}$ is the
density of saturated nuclear matter) consists of ultrarelativistic
highly degenerate electrons, very neutron rich nuclei and free
degenerate neutrons (dripped off the nuclei). With increasing
density in the inner crust, the fraction of free neutrons (among
bound and free nucleons) grows up, so that free neutrons give larger
contribution to the pressure, while atomic nuclei become loosely
bound as if dissolving in the neutron sea. Finally, the nuclei
disappear at $\rho_{\rm cc}$ which signals the transition to the
liquid core of the star. The crust as a whole is about 1 km thick
and contains about one percent of the neutron star mass.

The liquid core can extend to about $10\, \rho_0$. The outer core
($\rho_{\rm cc}\leq \rho \lesssim 2\,\rho_0$) consists of strongly
degenerate neutrons, with some admixture of protons, electrons and
muons. The inner core ($\rho \gtrsim 2 \, \rho_0$) may contain
hyperons, pion or kaon condensates or free quarks or the mixture of
these.  

The aim of this paper is to study a possible layer of {\it nuclear
mantle}  at the bottom of the inner crust containing various {\it
exotic nuclear clusters} called collectively {\it nuclear pasta}.
It was predicted by \citet{Ravenhalletal83} who investigated
properties of matter in core-collapsing supernovae. It has been
studied by many authors as summarized in a discussion below 
(also see \citealt{HPY07} for references to early
publications). A neutron star mantle can exist or not,
depending on specific properties of nuclear interactions at
subnuclear densities. If exist, it forms the layer between the
ordinary crust and the core, $\rho_{\rm cm} \leq \rho \leq \rho_{\rm
cc}$, where $\rho_{\rm cm}\approx 10^{14}$ g~cm$^{-3}$ is the outer
boundary of this layer. The mantle is often viewed as a part of the
inner crust. Although the mantle occurs in a narrow density
interval, its mass can be comparable to the mass of the entire
crust. As remarked above, the nuclei are loosely bound there. In
this case nuclear binding becomes as low as Coulomb energies of the
nuclei. Nuclear pasta forms when Coulomb energy is sufficiently
strong to rearrange nearly spherical atomic nuclei into exotic
nuclear clusters.

The studies of nuclear pasta can be divided into two parts. First,
one considers hot matter of subnuclear densities for core-collapsed
supernovae and proto-neutron stars where beta equilibrium is
absent. The second direction is to investigate colder pasta that can
be available in neutron stars under the beta equilibrium condition.
We are interested in the second possibility. Very roughly, a transition from the
first to the second case occurs at temperatures $T$ from about a few
times of $10^9$~K to $\sim 10^{10}$~K.

There are two major approaches to study nuclear pasta.

(i) First, one considers one-, two- or three
dimensional nuclear clusters using
different versions of the compressible liquid drop model (e.g.,
\citealt{Ravenhalletal83,HASHIMOTO84,LRP93,WIS00,WATANABE-IIDA03,
NAKAZATO09}) or Thomas-Fermi theories (e.g.,
\citealt{WK85,Oyamatsu93,OKAMOTO2013,GRILL2014}) with various mean
field interactions. The theory predicts four phases I--IV of nuclear
pasta in a neutron star mantle. The first phase I, which appears
with increasing density at $\rho>\rho_{\rm cm}$, is the phase of
rods (``spaghetti''). Nuclear structures become rod-like there,
immersed in a liquid of free neutrons. One studies neutron and
proton density profiles across these rods assuming that their length
is large (infinite). The phase of rods is followed by phase II of
slabs (``lasagna'' phase). The slabs are filled by nuclear matter
and immersed in neutron liquid. One calculates nuclear structure across the
slabs while the slabs themselves are assumed large
(infinite). The phase of slabs is followed by phase III of rod-like
bubbles (``anti-sphagetti''), where free neutrons fill infinitely
long bubbles aligned along certain axis, and the nuclear structures
occupy all other space. The final phase IV consists of nuclear
matter with spherical bubbles of neutron (and possibly proton)
liquid (``Swiss cheese''). Some calculations predict all four exotic
phases, while others predict fewer phases or no exotic phases at
all. Real rods or slabs are thought to be of finite size and occupy
some domains. The sizes of nuclear clusters and domains as well as
orientations of nuclear clusters in different domains are
uncertain.

(ii) Second, one simulates the composition and
dynamics of ensembles of nuclear clusters
with various versions of quantum molecular dynamics (e.g.,
\citealt{MARUYAMA98,KIDO2000,WATANABE2002,SCHUETRUMPF15}) or
classical molecular dynamics (e.g.,
\citealt{HOROWITZ04,HOROWITZ2005,DORSO2012,CAPLAN2014, DISORDER15}). Such
simulations predict a variety of possibilities, first of all nuclear
clusters of one type, e.g. rods, but of finite lengths and
different sizes. They predict also mixtures of different clusters
(e.g., rods and slabs). Recently, more exotic structures have been
obtained, such as slabs with a lattice of bubbles (``nuclear
waffles,'' \citealt{SCHNEIDER14}),  gyroids
\citep{NAKAZATO09,SCHUETRUMPF15}, intertwined lasagna and other
structures which do not resemble usual pasta \citep{ALCAIN14}, as
well as long lived topological defects (e.g., \citealt{DISORDER15}).
At high temperatures $T \gtrsim 10^{10}$ K nuclear pasta is mostly in liquid
state but at low temperatures (in neutron stars) it freezes into
ordered or disordered clusters.

The existence of nuclear pasta does not affect significantly the
equation of state of the matter and neutron star models. However it
can strongly influence neutrino emission (e.g.,
\citealt{LRP93,EBREMA1999,DURCA2004,HOROWITZ2005}) 
and transport properties of the neutron star mantle 
(e.g., \citealt{PONS2013,DISORDER15}). We
will focus on electron transport which is thought to be dominant
there. To simplify the problem we assume the existence of nuclear
pasta domains with nuclear clusters of one type in each domain. We
include the effects of strong magnetic fields which can be very
important.

\section{One domain}
\label{s:formalism}

\subsection{General expressions}
\label{s:general}

We start with the electron transport properties in one domain 
assuming that electron mean free paths are larger than 
spacing between nuclear clusters but smaller than domain size and
neglecting boundary effects. For example, consider ensembles of
rod-like or slab-like nuclear structures.  The electrons are mainly
scattered by proton charge distributions created by these
structures.

Consider a strongly degenerate and ultra-relativistic electron gas
which is slightly out of thermodynamic equilibrium due to the
presence of weak gradients of temperature $T$ and electron chemical
potential $\mu$ as well as due to a weak electric field $\bm{E}$.
Kinetic properties of the electrons are calculated from the
Boltzmann equation for the electron distribution function
$f(\bm{r},\bm{p},t)$,
\begin{equation}
   \frac{{\rm d}f}{{\rm d}t} \equiv \frac{\partial f}{\partial t}
   +\bm{v}\cdot\frac{\partial f}{\partial \bm{r}}
   -e\left(\bm{E}+\frac{\bm{v}}{c} \times \bm{B}  \right)\cdot
   \frac{\partial f}{\partial \bm{p}}=I\{f\},
\label{e:Boltzmann}
\end{equation}
where $\bm{v}$, $\bm{p}$ and $-e$ are the electron velocity,
momentum and electric charge, respectively; $\bm{B}$ is the external
magnetic field, and $I\{f\}$ is the collision integral. The
distribution function is normalized as
\begin{equation}
    n_{\rm e}(\bm{r},t)=\frac{2}{(2\pi \hbar)^3}\,
    \int {\rm d}^3\!\!p\,f(\bm{r},\bm{p},t),
\label{e:normaliz}
\end{equation}
$n_{\rm e}$ being the electron number density. Without any loss of
generality it is sufficient to consider a stationary problem, with
$\partial f/\partial t=0$. Because the electrons deviate from
equilibrium only slightly, we can set $f=f_0+\delta f$, where
\begin{equation}
    f_0=\left[\exp\left( \frac{\epsilon -\mu}{k_{\rm B}T}   \right)+1
    \right]^{-1}
\label{e:Fermi}
\end{equation}
is the local equilibrium Fermi distribution ($\epsilon$ being the
electron energy and $k_{\rm B}$ the Boltzmann constant), and $\delta
f$ is a small non-equilibrium correction.

It is well known (e.g., \citealt{KINETICS81}) that one can set
$f=f_0$ on the left-hand side of (\ref{e:Boltzmann}) in all the
terms except for the one containing the magnetic force (because the
latter term vanishes at $f=f_0$). Then
\begin{equation}
   \bm{v}\cdot\frac{\partial f_0}{\partial \bm{r}}
   -e\bm{E}\cdot
   \frac{\partial f_0}{\partial \bm{p}}
   -e\,\left(\frac{\bm{v}}{c} \times \bm{B}\right)\cdot
   \frac{\partial \delta f}{\partial \bm{p}}=I\{\delta f\},
\label{e:Boltzmann1}
\end{equation}
where $I\{\delta f\}$ is the linearized collision integral.

Since $f_0$ depends on $\bm{r}$ through $T(\bm{r})$ and
$\mu(\bm{r})$, equation (\ref{e:Boltzmann1}) can be rewritten as
\begin{equation}
   \bm{v}\cdot\bm{F}
   -e\left( \frac{\bm{v}}{c} \times \bm{B}\right)\cdot
   \frac{\partial \delta f}{\partial \bm{p}}=I\{\delta f\},
\label{e:Boltzmann2}
\end{equation}
where we have introduced the notations
\begin{equation}
  \bm{F}=\left( - \frac{\partial f_0}{\partial \epsilon} \right)\,
  \left(e \bm{E}_*+\frac{\epsilon - \mu}{T}\,\bm{\nabla}T  \right),
  \quad \bm{E}_*=\bm{E}+ \frac{\bm{\nabla}\mu}{e},
\label{e:FE*}
\end{equation}
$\bm{E}_*$ being the effective electric field. While calculating the
electron kinetic coefficients we can treat $\bm{\nabla}T$,
$\bm{\nabla} \mu$, $\bm{E}$ and $\bm{E}_*$ as arbitrary constant
small vectors.

At the next step we should specify the linearized collision integral
$I\{\delta f\}$. In order to study the main properties of kinetic
coefficients one can often use the traditional simplified relaxation
time approximation,
\begin{equation}
    I\{\delta f\}=-\frac{\delta f}{\tau}=-\nu\,\delta f,
\label{e:tau-nu}
\end{equation}
where $\tau$ is the relaxation time and $\nu=1/\tau$ is the
effective collision frequency. In the simplest case when electrons
scatter elastically by some heavy centers, one has $\nu=n_{\rm i}v
\sigma_{\rm tr}$, where $n_{\rm i}$ is the number density of scatterers
and $\sigma_{\rm tr}$ is the transport cross section; $\nu$, $\tau$
and $\sigma_{\rm tr}$ can depend on the electron energy $\epsilon$.
If one interests in a dipole-like deviation of the electron
distribution function from the equilibrium one (which is relevant
for studying electric and thermal conductivities as well as
thermopower), it is sufficient to set $\delta f= \bm{v}\cdot
\bm{u}$, where the vector $\bm{u}$ (to be determined) can depend on
the electron energy $\epsilon$ but not on orientation of its
momentum $\bm{p}$. Then $I\{\delta f\}=-\nu\, \bm{v} \cdot \bm{u}$.

The case of nuclear pasta is more complicated because we deal with
anisotropic medium. As far as nuclear clusters are concerned, we
assume that we have a symmetry plane $(x,y)$ (perpendicular to
nuclear rods or plates) and a symmetry axis $z$ (perpendicular to
the plane).  Without any loss of generality we may assume that the
magnetic field $\bm{B}$ lies in the $(x,z)$ plane.

A natural generalization of the relaxation time approximation
(\ref{e:tau-nu}) to this case is
\begin{equation}
    I\{\delta f\}=-\nua \,v_z u_z-\nup \, \bm{v}_{\rm p}
     \cdot \bm{u}_{\rm p},
\label{e:collision_integral}
\end{equation}
where $\nua$ and $\nup$ are generally different effective collision
frequencies along the symmetry axis and in the symmetry plane,
respectively; $\bm{v}_{\rm p}=(v_x,v_y)$; $\bm{u}_{\rm
p}=(u_x,u_y)$, and $\bm{u}$, as before, can depend on $\epsilon$ but
not on orientation of $\bm{p}$. The difference between $\nua$ and
$\nup$ reflects the anisotropy of electron collision frequencies in
non-spherical nuclear clusters. Such an approximation is often
used for studying electron transport in anisotropic solids (see,
e.g., \citealt{ASKEROV94}). In the absence of anisotropy, one has
$\nua =\nup$ and the problem reduces to a single traditional
effective scattering frequency (\ref{e:tau-nu}) in isotropic medium.

\renewcommand{\arraystretch}{1.2}
\begin{table}
\caption{Possible pasta phases in a neutron
star mantle}
\begin{center}
\begin{tabular}{ c l  c c }
\toprule
Number & Phase & Leading $\nu$ & Weaker $\nu$  \\
\midrule
I & Rods    &  $\nup$ &  $\nua$    \\
II & Slabs &   $\nua$ & $\nup$    \\
III & Rod-like bubbles  &     $\nup$ & $\nua$   \\
IV & Spherical bubbles &  $\nua= \nup$ & --- \\
 \hline
\end{tabular}
\label{tab:pasta}
\end{center}
\end{table}

As discussed in Section \ref{sec:intro}, there could be four phases
I--IV of nuclear pasta in a neutron star mantle; they are listed in
Table \ref{tab:pasta} in order of appearance with increasing density
at $\rho>\rho_{\rm cm}$.   Phase I consists of rods. Form symmetry
consideration one can expect that the most frequent electron
momentum transfer occurs when electrons move across the symmetry
axis (across the rods), so that $\nup$ should be larger than $\nua$.
We will call $\nup$ the leading collision frequency and $\nua$ the
weaker collision frequency (in phase I).

Phase I is followed by phase II of slabs (Table \ref{tab:pasta}).
Now the leading collision frequency $\nua$ corresponds to electrons
moving along the symmetry axis; they can scatter effectively by
slabs which are perpendicular to the axis. The weaker collision
frequency is then $\nup$.

After slabs one has phase III of rod-like bubbles, where
the collision frequencies are expected to have the same
ordering as in phase I, $\nup \gg \nua$.

The final phase IV is that of spherical bubbles. In this case
$\nua=\nup$ (electron collisions are isotropic) as in the case of
ordinary spherical nuclei in the main body of the neutron star crust
(at $\rho<\rho_{\rm cm}$).

For example, according to model I of \citet{Oyamatsu93}, $\rho_{\rm
mc}=0.973 \times 10^{14}$ g~cm$^{-3}$. The transition to phase II
occurs at $\rho=1.24 \times 10^{14}$ g~cm$^{-3}$; to phase III at
$\rho=1.37 \times 10^{14}$ g~cm$^{-3}$, and to phase IV at
$\rho=1.42 \times 10^{14}$ g~cm$^{-3}$. Finally, phase IV ends at
$\rho_{\rm cc}=1.43 \times 10^{14}$ g~cm$^{-3}$. In a typical
neutron star of  mass $1.4 \,{\rm M}_\odot$ and radius about 12 km
the nuclear pasta layer can be about 100 m thick and have mass $\sim
0.01\, {\rm M}_\odot$; it is nearly as massive as the entire
overlying ordinary crust of spherical nuclei.

Equation (\ref{e:collision_integral}) reduces the linearized
Boltzmann equation (\ref{e:Boltzmann2}) to
\begin{equation}
   \bm{v}\cdot\bm{F}
   -\omega \,(\bm{v}\times \bm{b})\cdot \bm{u}=
   -\nua \,v_z u_z-\nup \, \bm{v}_{\rm p}
     \cdot \bm{u}_{\rm p},
\label{e:Boltzmann3}
\end{equation}
where $\omega=eBc/\epsilon$ is the electron gyrofrequency
($\epsilon$ includes the electron rest-mass energy), and
$\bm{b}=\bm{B}/B$ is the unit vector in the direction of $\bm{B}$.
In the matrix form
\begin{equation}
  \left(
  \begin{array}{ccc}
   -\nup & -\omega b_z & 0 \\
   \omega b_z & -\nup & -\omega b_x \\
   0     & \omega b_x  & -\nua
  \end{array}
  \right)\,
  \left(
  \begin{array}{c}
   u_x \\ u_y \\ u_z
  \end{array}
  \right)=
  \left(
  \begin{array}{c}
   F_x \\ F_y \\ F_z
  \end{array}
  \right).
\label{e:matrix}
\end{equation}
The solution can be written as
\begin{equation}
     u_\alpha=-\frac{1}{\Delta}\,A_{\alpha \beta}F_\beta,
\label{e:vector_u}
\end{equation}
where $\Delta=\nup^2 \nua + \omega^2 (\nup b_x^2+\nua b_z^2)$. The
matrix $A_{\alpha \beta}$, which will be denoted as $\hat{A}$, is
\begin{equation}
    \hat{A}=
\left(
  \begin{array}{ccc}
   \nua\nup+\omega^2 b_x^2 & -\omega b_z \nua &
   \omega^2 b_x b_z \\
   \omega b_z \nua & \nua \nup & -\omega b_x \nup \\
   \omega^2 b_x b_z      & \omega b_x \nup & \nup^2+\omega^2
   b_z^2
  \end{array}
  \right),
\label{e:matrixA}
\end{equation}
where $b_z=\cos \theta$, $b_x=\sin \theta$ and $\theta$ is the angle
between $\bm{B}$ and the symmetry axis $z$.

This gives us the vector $\bm{u}$ and
$f=f_0+\delta f$. Then we can calculate the densities of electric
and thermal currents,
\begin{equation}
  \bm{j}=-\frac{2e}{(2 \pi \hbar)^3} \,\int {\rm d}^{3}\!p\;\bm{v}\,f, \quad
  \bm{q}=\frac{2}{(2 \pi \hbar)^3} \,\int {\rm d}^3\!p\;\bm{v}\,(\epsilon-\mu)\,f.
\label{e:jq}
\end{equation}
Formally, $\bm{j}$ and $\bm{q}$ have the following structure
\begin{equation}
   \bm{j}=\hat{\sigma}\,\bm{E}_*+\hat{\beta}\,\bm{\nabla}T,
    \quad
     \bm{j}=-\hat{\beta}T\,\bm{E}_*-\hat{\eta}\,\bm{\nabla}T,
\label{e:kin-coeff}
\end{equation}
where hat means $3 \times 3$ matrix (tensor) of anisotropic
kinetic coefficients, $\hat{\sigma}$ is the electric conductivity
tensor, while two other tensors, $\hat{\beta}$ and $\hat{\eta}$, are
auxiliary. The same tensor $\hat{\beta}$ enters both expressions,
for $\bm{j}$ and $\bm{q}$, as a consequence of the Onsager
reciprocal relations.

For applications, it is instructive to rewrite (\ref{e:kin-coeff})
in the form
\begin{equation}
   \bm{E}_* = \hat{\cal R} \, \bm{j}-\hat{\alpha} \,\bm{\nabla}T,
    \quad
     \bm{q}=-\hat{\alpha}T \bm{j}-\hat{\kappa} \,\bm{\nabla}T,
\label{e:kin-coeff1}
\end{equation}
where $\hat{\cal R}=\hat{\sigma}^{-1}$, $\hat{\kappa}$ and
$\hat{\alpha}$ are the tensors of electric resistivity, thermal
conductivity and thermopower, respectively. Again, the same tensor
$\hat{\alpha}$ enters the expressions for $\bm{E}_*$ and $\bm{q}$
owing to the Onsager relations. The thermal conductivity and
thermopower are given by $\hat{\kappa}=\hat{\eta}-T
\hat{\beta}\,\hat{\cal R}\,\hat{\beta}$ and $\hat{\alpha}=\hat{\cal
R}\,\hat{\beta}$, respectively.

So far our consideration has been quite general, valid for the
electron gas of any relativity and degeneracy. The analysis is
considerably simplified for strongly degenerate electrons. In this
case the main contribution to electron transport comes from the
electrons with energies $|\epsilon -\mu| \lesssim k_{\rm B}T$ 
in the narrow thermal width of the Fermi level. Standard
calculations yield
\begin{equation}
  \hat{\sigma}=\frac{e^2 n_{\rm e} \hat{A}}{ m_* \Delta},\quad
     \hat{\kappa}\approx \hat{\eta}
     =\frac{\pi^2 n_{\rm e} k_{\rm B}^2 T\hat{A}}{3  m_*\Delta},
\label{e:sigma,kappa}
\end{equation}
where $\hat{A}$ is given by (\ref{e:matrixA}), $\Delta$ is defined
in (\ref{e:vector_u}), $m_*=\mu/c^2$ is the effective electron mass
at the Fermi surface ($\mu$ includes $m_{\rm e}c^2$), while
$\omega$, $\nua$ and $\nup$ have to be taken at $\epsilon=\mu$. Note
that $\hat{\sigma}$ and $\hat{\kappa}$ satisfy the Wiedemann-Franz
law,
\begin{equation}
\hat{\kappa}=\frac{\pi^2 k_{\rm B}^2 T}{3e^2}\, \hat{\sigma},
\label{e:WF}
\end{equation}
which is natural for strongly degenerate electrons in the
relaxation-time approximation (e.g. \citealt{ZIMAN60}).

The tensors $\hat{\sigma}$ and $\hat{\kappa}$ have similar
structure, determined by the matrix $\hat{A}$. Therefore, either
electric or thermal conduction is defined by {\it six} different
transport coefficients of {\em three} types:
\begin{enumerate}

\item Three diagonal terms ($A_{xx}$, $A_{yy}$, $A_{zz}$) are generally
different reflecting anisotropy of the transport along all three axes;

\item There are two basically different
Hall coefficients ($A_{xy}=-A_{yx}$; $A_{yz}=-A_{zy}$) and two different
Hall parameters,
\begin{equation}
  x_{\rm a}=\omega/\nua, \quad
  x_{\rm p}=\omega/\nup,
\label{eq:hall-param}
\end{equation}
associated with
two collision frequencies, $\nua$ and $\nup$; they describe
gyrotropic character of electron transport in a magnetic field;

\item There are two equal coefficients $A_{xz}=A_{zx}$ for non-vanishing
$B_x$ and $B_z$;

\item One has $A_{\alpha \beta}(\bm{B})=A_{\beta \alpha}(-\bm{B})$, in agreement
with Onsager relations.

\end{enumerate}
Naturally, the electron transport in nuclear pasta is richer in physics than
in ordinary neutron star matter.

Let us present also the resistivity tensor
\begin{equation}
   \hat{\cal R}= \frac{m_*}{e^2n_{\rm e}}\;
     \left(
  \begin{array}{ccc}
   \nup & \omega b_z & 0 \\
   -\omega b_z & \nup & \omega b_x \\
   0     & -\omega b_x  & \nua
  \end{array}
  \right).
\label{e:hatR}
\end{equation}
Its structure is simpler than that of $\hat{\sigma}$ or $\hat{\kappa}$.

If $\hat{\sigma}$ is found from equation (\ref{e:sigma,kappa}), one
can calculate the thermopower tensor as
\begin{equation}
   \hat{\alpha}=\frac{\pi^2 k_{\rm B}^2 T \hat{\cal R}}{3e}\,
    \frac{\partial \hat{\sigma}(\mu)}{\partial \mu}.
\label{e:thermopower}
\end{equation}
Such relations are standard \citep{ZIMAN60} for scalar thermopower
of strongly degenerate electrons in the relaxation time
approximation. Evidently, they remain valid for tensor quantities. 
In order to use them one should express
$\omega$, $\nua$, and $\nup$ in $\hat{\sigma}$ as functions of
$\epsilon$ and set $\epsilon=\mu$. In addition, one should express
$n_{\rm e}$ through $\mu$ and differentiate over $\mu$ afterwards.

The presented expressions are sufficient to estimate the electron
transport properties in one domain of nuclear pasta. Below we
analyze the transport coefficients in limiting cases. Since the
structure of $\hat{\sigma}$ and $\hat{\kappa}$ is the same, we will
mainly discuss $\hat{\sigma}$ meaning that one can immediately find
$\hat{\kappa}$ using equation (\ref{e:WF}).

\subsection{No magnetic field}
\label{s:B=0}

In the limit of ${\bm B} \to 0$ the transport coefficients are
greatly simplified,
\begin{equation}
   \hat{\sigma}
     = \frac{e^2 n_{\rm e}}{m_*}
     \left(
     \begin{array}{ccc}
      \nup^{-1} & 0 & 0 \\
        0 & \nup^{-1} & 0 \\
        0 & 0 & \nua^{-1}
     \end{array}
    \right).
\label{e:B=0}
\end{equation}
Here, the conductivity is characterized by the two different kinetic
coefficients, along and across the symmetry axis,
\begin{equation}
  \sigma_{\rm a0}= \frac{e^2 n_{\rm e}}{\nua m_*}, \quad
    \sigma_{\rm p0}= \frac{e^2 n_{\rm e}}{\nup m_*}.
\label{e:B=0,sigmaap}
\end{equation}
In phases I and III of nuclear rods or rod-like bubbles one expects
$\nup \gg \nua$ (Table \ref{tab:pasta}). Then the conductivity
$\sigma_{\rm a0}$ along the symmetry axis is much larger than the
leading conductivity $\sigma_{\rm p0}$ across this axis. If we view
the leading transport as `normal,' then the transport in the
perpendicular direction is abnormally fast. In phase II of nuclear
slabs one has $\nua \gg \nup$. The normal leading transport along
the symmetry axis is much slower than across it ($\sigma_{\rm a0}\ll
\sigma_{\rm p0}$). Finally, in phase IV of spherical bubbles $\nua =
\nup$, so that the transport is isotropic (`normal' in all
directions, as in ordinary neutron star crust).

\subsection{Finite magnetic field along or across symmetry axis}
\label{s:Bneq0}

If $\bm{B} \neq 0$, transport properties are generally complicated. The
simplest case takes place if $\bm{B}$ is directed along the symmetry axis.
Then
\begin{equation}
   \hat{\sigma}
     = \frac{e^2 n_{\rm e}}{m_*\nup(1+x_{\rm p}^2)}
     \left(
     \begin{array}{ccc}
     1  & -{x_{\rm p}} & 0 \\
        {x_{\rm p}} & 1 & 0 \\
        0 & 0 & {\nup}(1+x_{\rm p}^2)/\nua
     \end{array}
    \right).
\label{e:B1}
\end{equation}
The conductivity in the symmetry plane is gyrotropic (governed
by the magnetic field and the collision frequency $\nup$), while the
conductivity along the symmetry axis is unaffected by $\bm{B}$ and
determined by $\nua$.

If $\bm{B}$ lies in the symmetry plane,  along the $x$ axis, then
\begin{equation}
    \hat{\sigma}
      = \frac{e^2 n_{\rm e}}{m_*\nup(\nua \nup+\omega^2)}
     \left(
     \begin{array}{ccc}
     \nua \nup+\omega^2  & 0 & 0 \\
        0 & \nua \nup & -\omega \nup \\
        0 & \omega \nup & \nup^2
     \end{array}
    \right).
\label{e:B2}
\end{equation}
The conductivity along $\bm{B}$ is again unaffected by the magnetic
field (being naturally determined by $\nup$), while the conductivity
in the $yz$ plane (across $\bm{B}$) is gyrotropic and regulated by
$\nup$ and $\nua$.

If $\nua=\nup=\nu_0$, the conductivity is the same as in a basically
isotropic medium immersed in a magnetic field; 
see Section \ref{s:spherical-bubbles}.

\begin{figure*}
\includegraphics[width=0.33\textwidth]{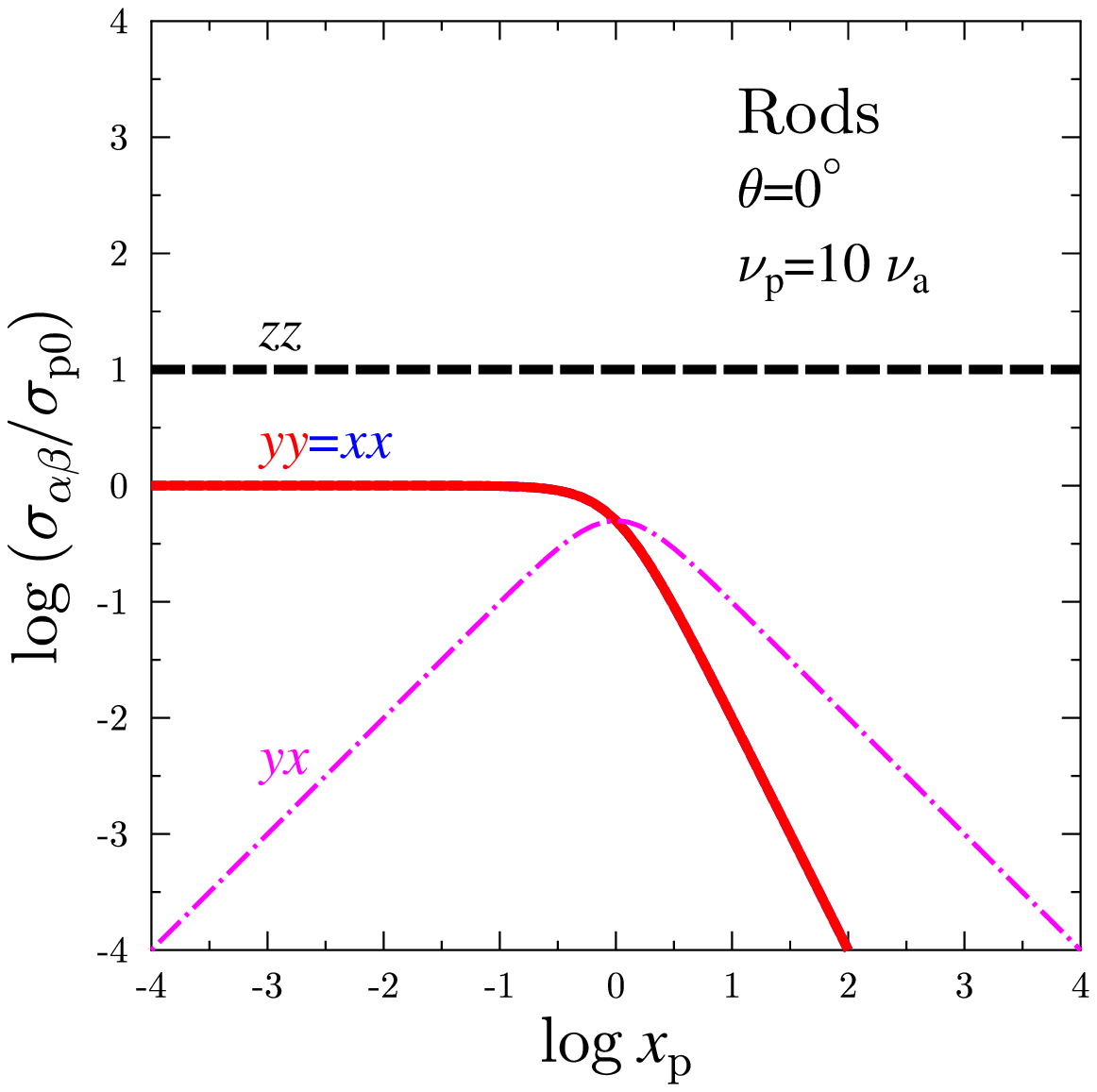}%
\includegraphics[width=0.33\textwidth]{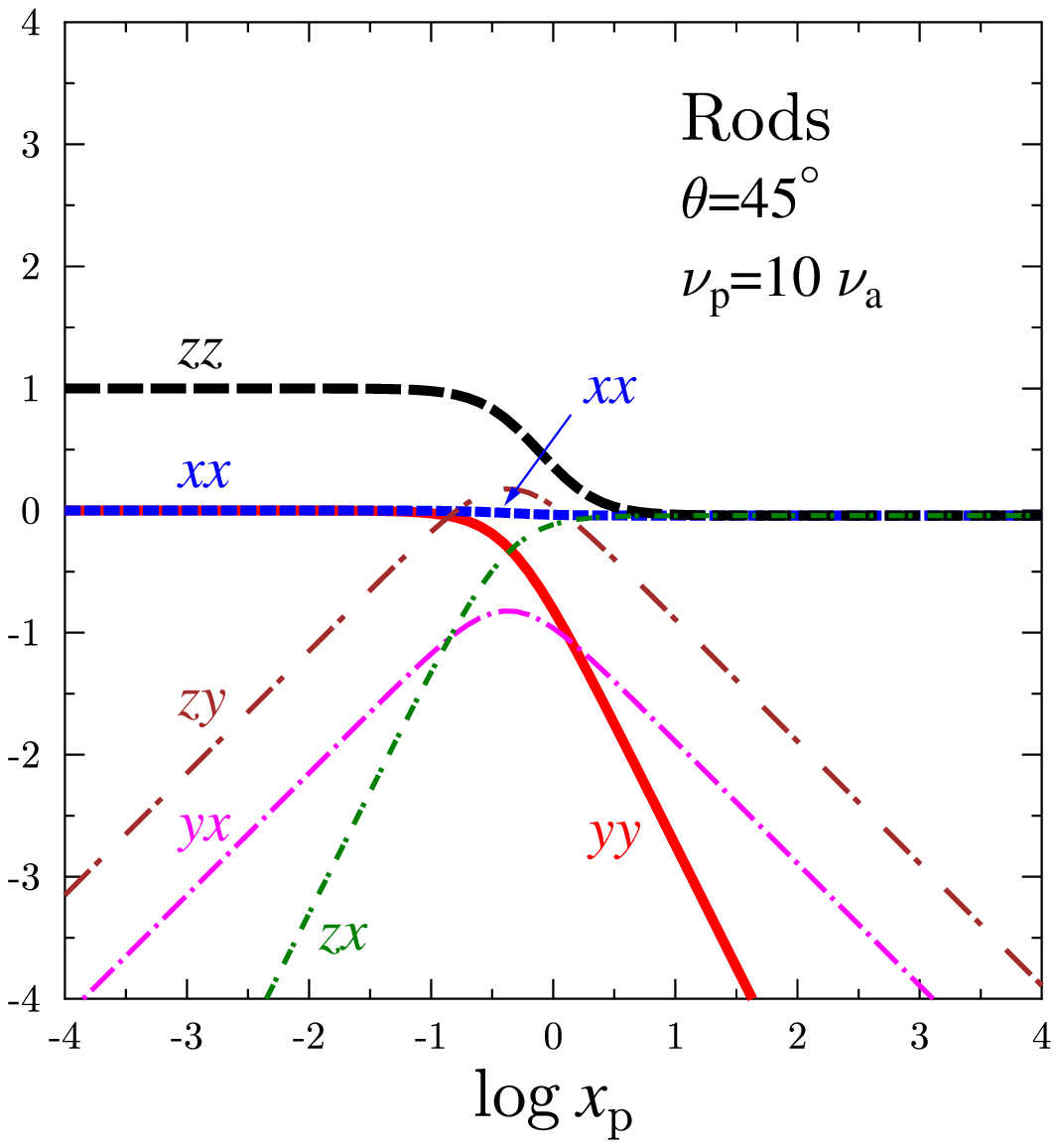}%
\includegraphics[width=0.33\textwidth]{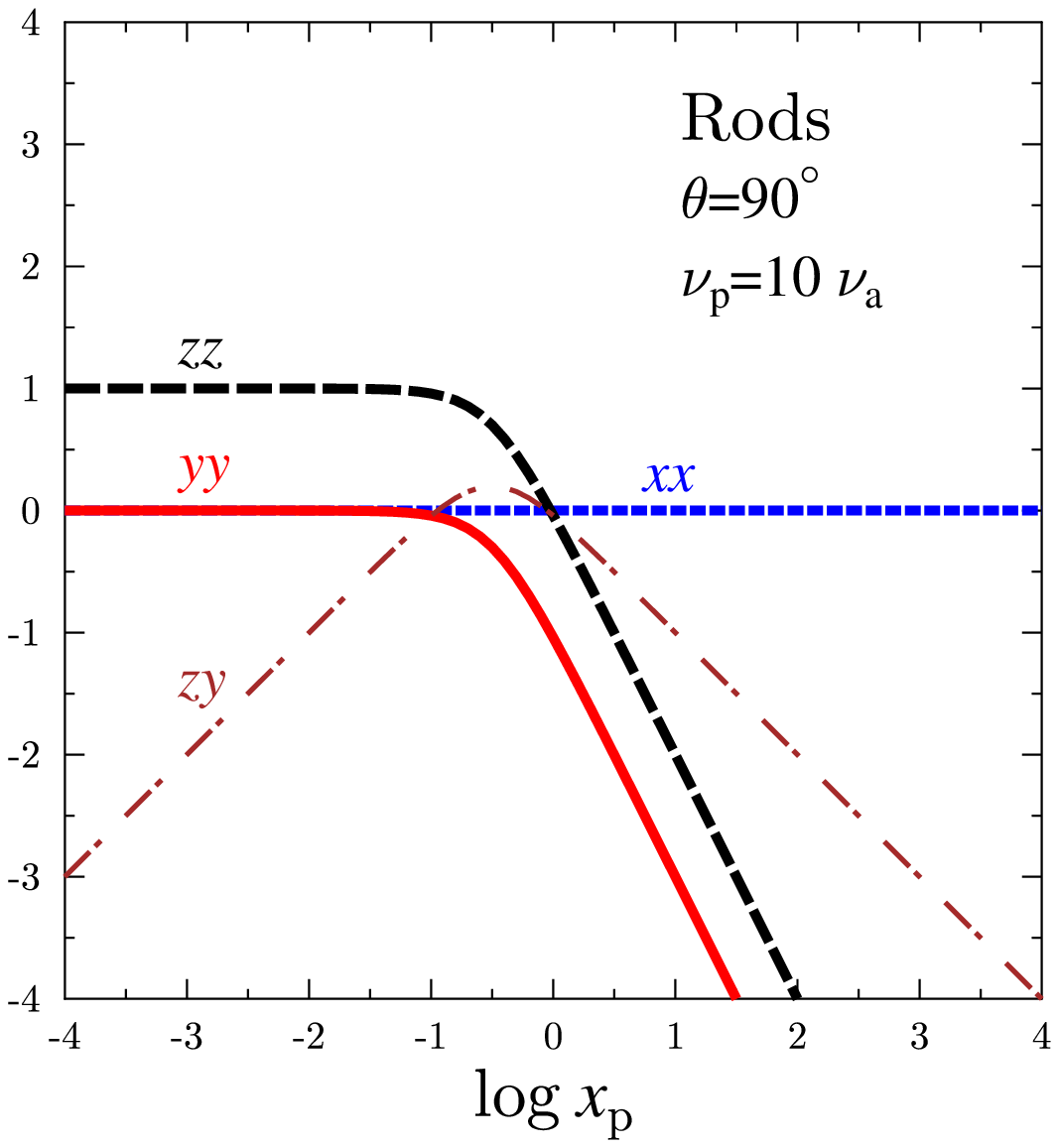}
\caption{(Color online) Components of the electron
conductivity tensor $\sigma_{\alpha \beta}$  in phases I or III (in
units of the leading conductivity $\sigma_{\rm p0}$ across
the symmetry axis for $\bm{B}=0$) versus the leading Hall parameter
$x_{\rm b}=\omega /\nu_{\rm b} \propto B$
at three angles $\theta=0$, 45 and $90^\circ$ (from left
to right) between $\bm{B}$ and the symmetry axis. The ratio of the
effective electron collision frequencies across and along the
symmetry axis is taken to be $\nup/\nua=10$. Shown are
all six generally different tensor components, $\sigma_{xx}$, $\sigma_{yy}$,
$\sigma_{zz}$ (thick lines of different types) and $\sigma_{yx}$,
$\sigma_{zx}$ and $\sigma_{zy}$ (thin lines).} \label{fig:rods}
\end{figure*}

\begin{figure*}
\includegraphics[width=0.33\textwidth]{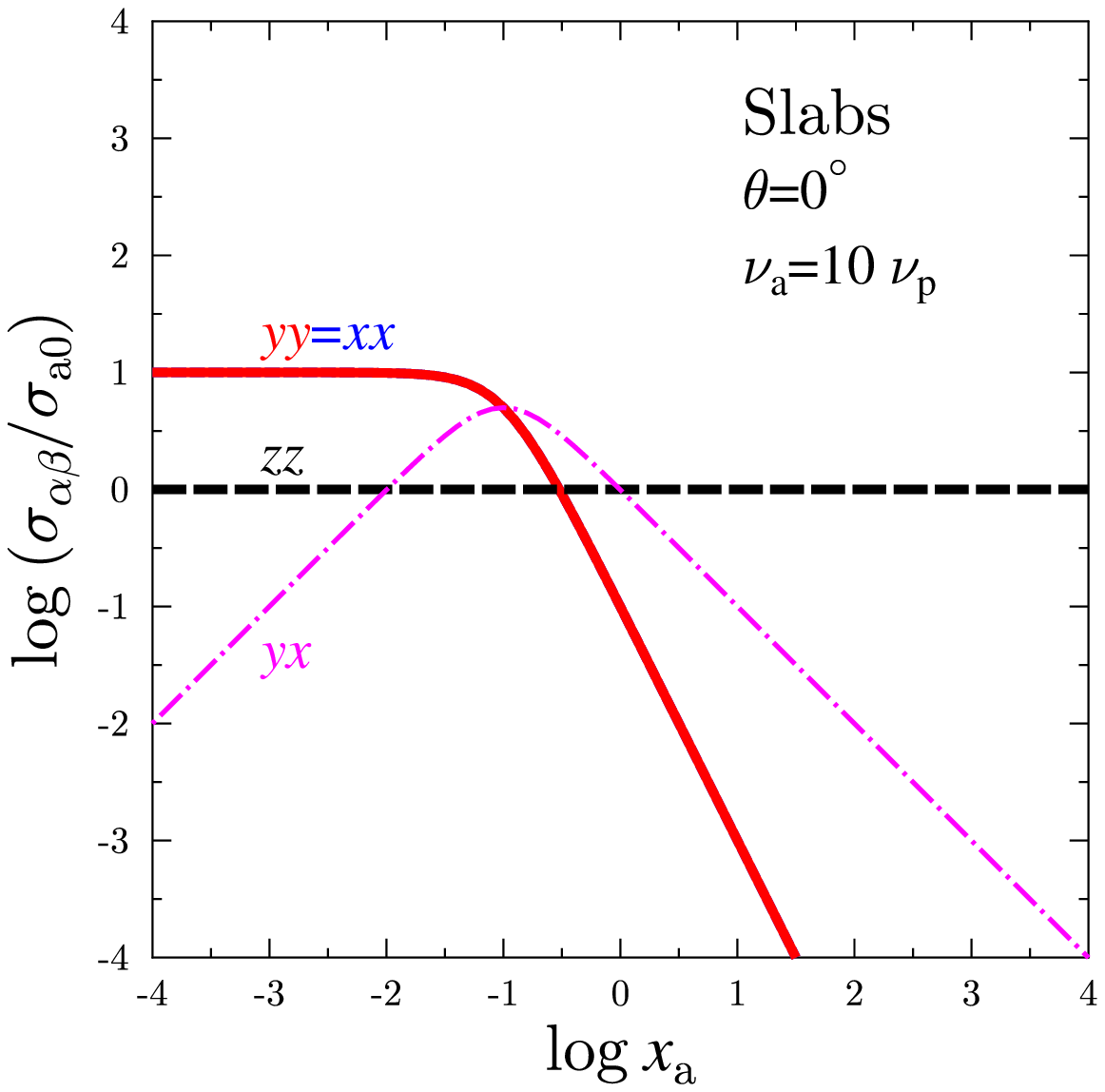}%
\includegraphics[width=0.33\textwidth]{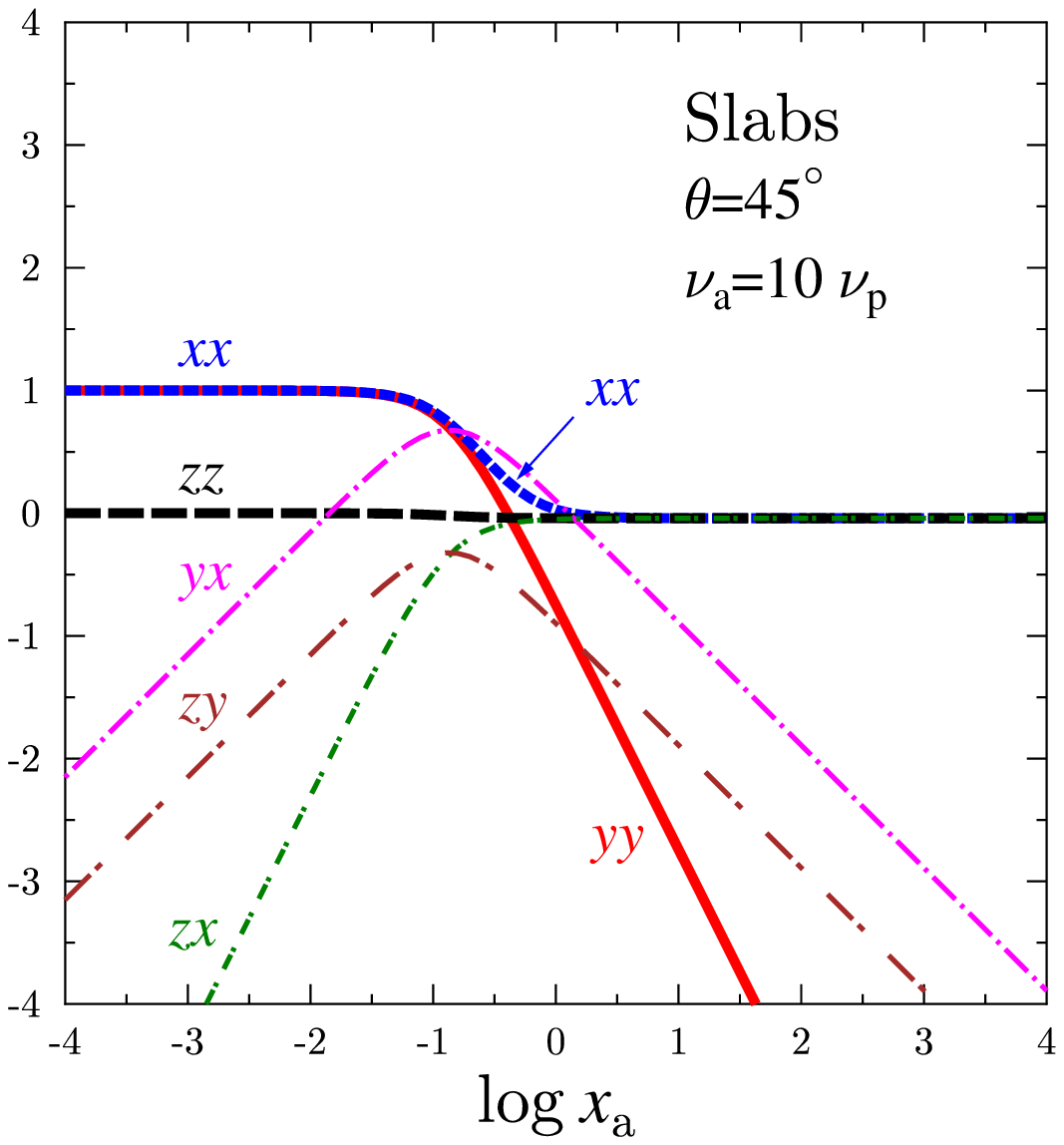}%
\includegraphics[width=0.33\textwidth]{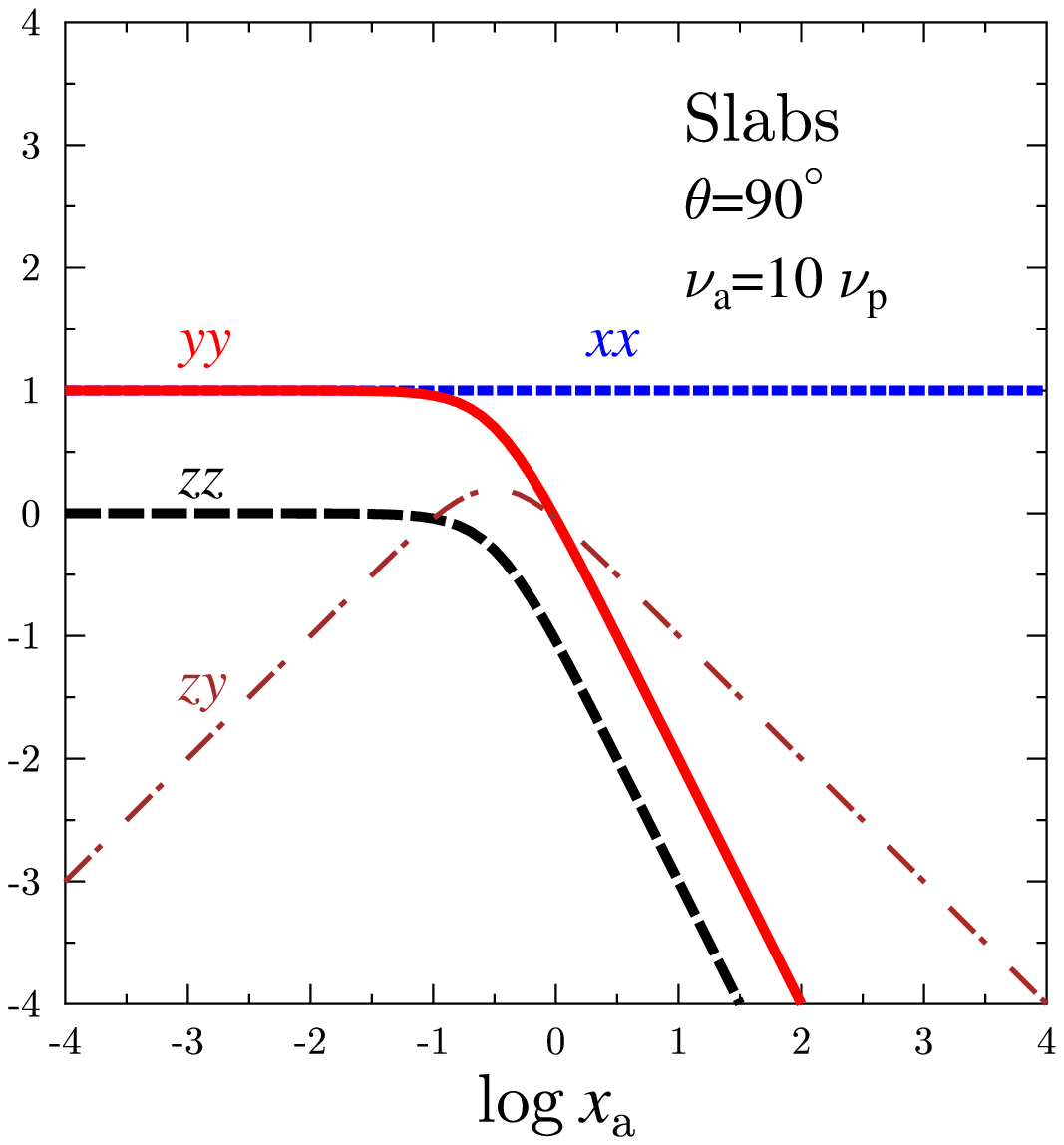}
\caption{(Color online) Same as in Fig.\ \ref{fig:rods} for electron
conductivity tensor $\sigma_{\alpha \beta}$ in phase II (in units of
the leading conductivity $\sigma_{\rm a0}$ along the
symmetry axis for $\bm{B}=0$) versus the leading Hall parameter
$x_{\rm a}=\omega /\nua$ (proportional to $B$) at $\theta=0$,
45 and $90^\circ$ (from left to right) and $\nua/\nu_{\rm
p}=10$.} \label{fig:slabs}
\end{figure*}

\subsection{Three regimes}
\label{sec:3regimes}

If we increase $B$ from $B=0$, the Hall parameters
$x_{\rm p}$ and $x_{\rm a}$ grow up and the electron transport is
affected by the magnetic field. The regime of sufficiently small
(smaller than 1) Hall
parameters corresponds to weakly magnetized electrons, whereas the
regime of high (higher than 1) Hall parameters  to strongly
magnetized electrons. Because $x_{\rm p}$ and $x_{\rm a}$ are not
equal, there is also the regime of intermediate magnetization in
which the electrons are already magnetized with respect to one
Hall parameter but are still not magnetized with respect to the
other one. These three magnetization regimes have different effects
on various conductivity components at different inclinations
$\theta$ of $\bm{B}$ to the symmetry axis.

As for the $B$-dependence of the conductivity, all conductivity
components are divided into non-Hall ($xx$,
$yy$, $zz$, and $zx$ ones) and Hall ones ($yx$ and $zy$; we do not mention
associated non-diagonal components like $xz$ or $xy$). The
non-Hall components are even functions of $\omega \propto B$
while the Hall components are odd functions. In
the limit of weak electron magnetization the diagonal components
stay nearly constant, while $\sigma_{zx}$ increases as $B^2$.
All non-vanishing Hall components increase proportionally to $B$.

In the limit of high electron magnetization ($x_{\rm a}\gg 1$,
$x_{\rm p} \gg 1$) we obtain
\begin{eqnarray}
    \hat{\sigma}&
      = & \frac{e^2 n_{\rm e}}{m_*(\nup \, \sin^2 \theta +
            \nua \cos^2 \theta)}
\nonumber \\
  & \times &   \left(
     \begin{array}{ccc}
     \sin^2\theta +(x_{\rm a} x_{\rm p})^{-1}  &
        - x_{\rm a}^{-1} \cos \theta  & \cos \theta \, \sin \theta \\
      x_{\rm a}^{-1} \cos \theta   & (x_{\rm a} x_{\rm p})^{-1} &
            -x_{\rm p}^{-1} \sin \theta \\
      \cos \theta \, \sin \theta   & x_{\rm p}^{-1} \sin \theta &
             \cos^2\theta + x_{\rm p}^{-2}.
     \end{array}
    \right).
\label{e:highB}
\end{eqnarray}
With increasing $B$ in this regime the non-Hall components either
stay constant (as $\sigma_{xx}$, $\sigma_{zz}$ and $\sigma_{zx}$) or
decrease proportionally to $B^{-2}$, whereas the Hall components
decrease as $B^{-1}$. The decrease of the conductivity components at
strong magnetization is evidently explained by frequent
rotation of electrons around magnetic field lines, with many
rotations between successive collisions.

\subsection{Resistivity}
\label{sec:resist}

The structure of the electric resistivity tensor $\hat{\cal R}$,
equation (\ref{e:hatR}), is simpler than of the conductivity. The
diagonal components ${\cal R}_{xx}={\cal R}_{yy}\equiv {\cal R}_{\rm
p}$ and ${\cal R}_{zz}={\cal R}_{\rm a}$ are totally independent of
$\bm{B}$, being equal to corresponding inverse diagonal components
(\ref{e:B=0,sigmaap}) of the electric conductivity tensor at $B=0$,
${\cal R}_{\rm p}=1/\sigma_{\rm p0}$ and  ${\cal R}_{\rm
a}=1/\sigma_{\rm a0}$. The leading resistivity (${\cal R}_{\rm p}$
or ${\cal R}_{\rm a}$) is higher than the resistivity in the
opposite direction. The Hall components ${\cal R}_{xy}=-{\cal
R}_{yx}$ and ${\cal R}_{yz}=-{\cal R}_{zy}$ are directly
proportional to $\omega \propto B$. In the regime of weak
magnetization the diagonal components are larger than the Hall ones
while in the the regime of strong magnetization the Hall components
are larger. Actually, there is only one non-trivial Hall component
of $\hat{\cal R}$ instead of two formal Hall components ${\cal
R}_{xy}$ and ${\cal R}_{yz}$ in (\ref{e:hatR}). To prove this one
can rotate coordinate axes in the $xz$ plane in such a way for the
$z$-axis to become parallel to $\bm{B}$. After this rotation the
tensor would contain the only one Hall component ${\cal
R}_{xy}=-{\cal R}_{yx}= m_*\omega/(e^2 n_{\rm e})$ totally
independent of electron collision frequencies $\nup$ and $\nua$
(proving that the Hall components do not produce any dissipation
although produce an important effect of Hall drift of the magnetic
field). Such a rotation trick does not work out for conductivity
tensors -- there are generally two non-trivial Hall components of
the conductivity.

The simplicity of the electric resistivity tensor is also pronounced
in the expression for the Joule heat $Q=(\bm{j}\cdot \hat{\cal R}
\cdot \bm{j})$ [erg~s$^{-1}$~cm$^{-3}$],
\begin{equation}
    Q={\cal R}_{\rm p} j_{\rm p}^2+{\cal R}_{\rm a} j_{\rm a}^2,
\label{e:joule}
\end{equation}
where $\bm{j}_{\rm p}$ and
$j_{\rm a}$ are the components of the electron current $\bm{j}$
across and along the symmetry axis, respectively.

\subsection{Phase of spherical bubbles}
\label{s:spherical-bubbles}

This phase is the simplest one because the medium is basically
isotropic, $\nua=\nup=\nu_0$. The conductivity and resistivity
tensors are especially simple in the coordinate system with the
$z$-axis parallel to $\bm{B}$. For instance, the electric
conductivity and resistivity tensors take the well known form
\begin{equation}
   \hat{\sigma}=
     \left(
     \begin{array}{ccc}
      \sigma_\perp & -\sigma_{\rm H} & 0 \\
        \sigma_{\rm H} & \sigma_\perp & 0 \\
        0   & 0 & \sigma_\parallel
     \end{array}
     \right),\quad
 \hat{\cal R}=
     \left(
     \begin{array}{ccc}
      {\cal R}_\perp & {\cal R}_{\rm H} & 0 \\
        -{\cal R}_{\rm H} & {\cal R}_\perp & 0 \\
        0   & 0 & {\cal R}_\parallel
     \end{array}
     \right),
\label{e:sph}
\end{equation}
where the subscripts $\parallel$ and $\perp$ refer to
conductivities and resistivities
parallel and perpendicular to $\bm{B}$, respectively, while
the subscript H labels the Hall terms. The Hall
terms in $\hat{\sigma}$ and $\hat{\cal R}$ have naturally different
signs. The expressions
for these components are
\begin{equation}
 \left(
  \begin{array}{c}
    \sigma_\parallel \\
    \sigma_\perp \\
    \sigma_{\rm H}
    \end{array}
    \right)
 =
 \sigma_0
 \left(
 \begin{array}{c}
  1 \\
    1/(1+x^2)\\
    x/(1+x^2)
 \end{array}
 \right),
\quad
\left(
  \begin{array}{c}
    {\cal R}_\parallel \\
    {\cal R}_\perp \\
    {\cal R}_{\rm H}
    \end{array}
    \right)
 = \frac{1}{\sigma_0}
 \left(
 \begin{array}{c}
  1 \\
    1 \\
    x
 \end{array}
 \right),
\label{e:sph1}
\end{equation}
with $\sigma_0={e^2 n_{\rm e}}/{\nu_0 m_*}$,
$x=\omega /\nu_0$ being the single Hall parameter in the
given case.
Therefore, $\sigma_\parallel=\sigma_0=1/{\cal R}_\parallel=1/{\cal R}_\perp$
are independent of $\bm{B}$, and ${\cal R}_{\rm H} \propto B$.
In the regime of weak magnetization, at $x\ll 1$, one has  nearly
constant  $\sigma_\perp \approx \sigma_\parallel$
but $\sigma_{\rm H}\propto B$, while at $x \gg 1$ one has
$\sigma_\perp \propto B^{-2}$ and $\sigma_{\rm H} \propto \perp B^{-1}$.

\begin{figure*}
\includegraphics[width=0.33\textwidth]{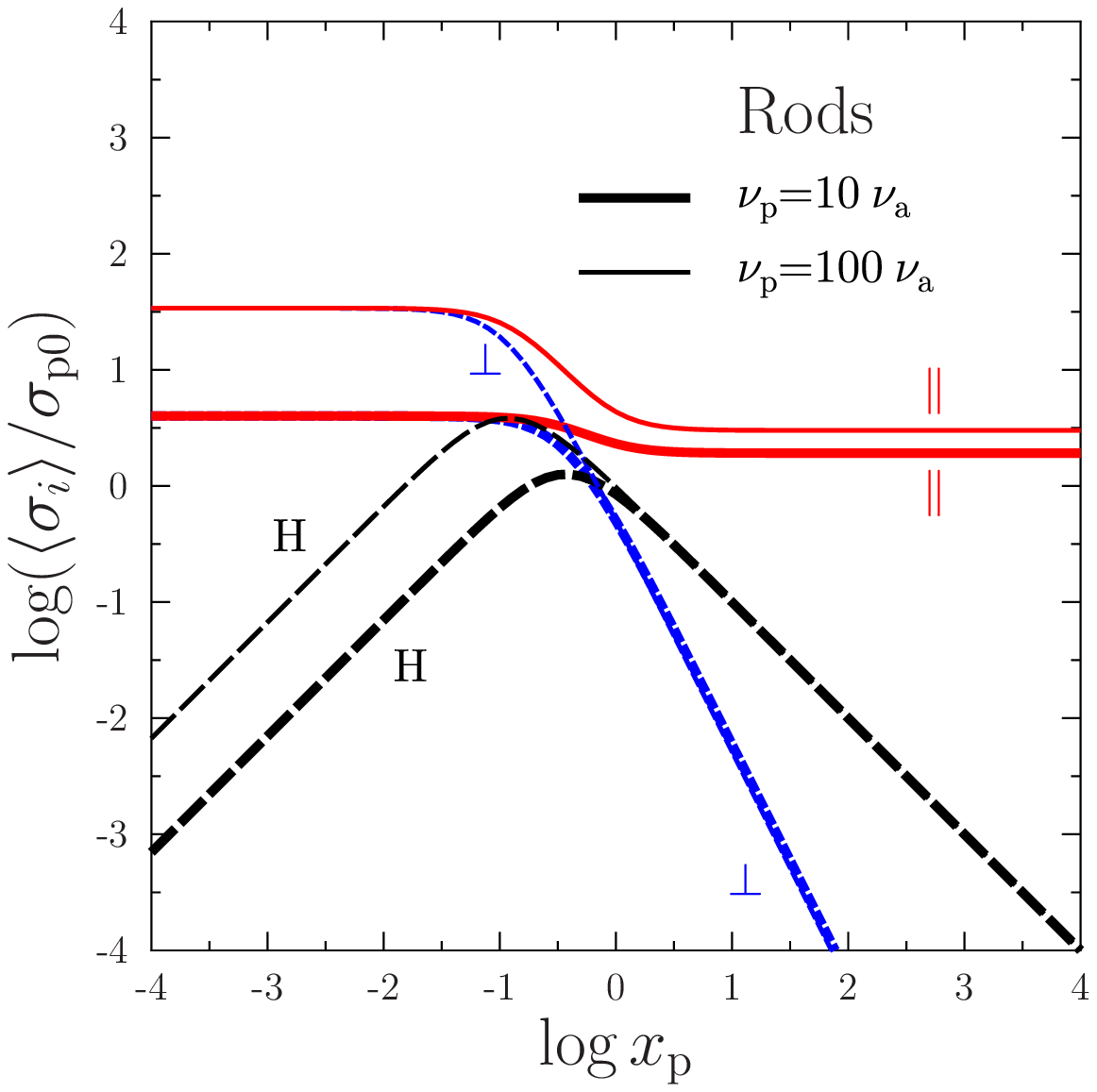}%
\includegraphics[width=0.33\textwidth]{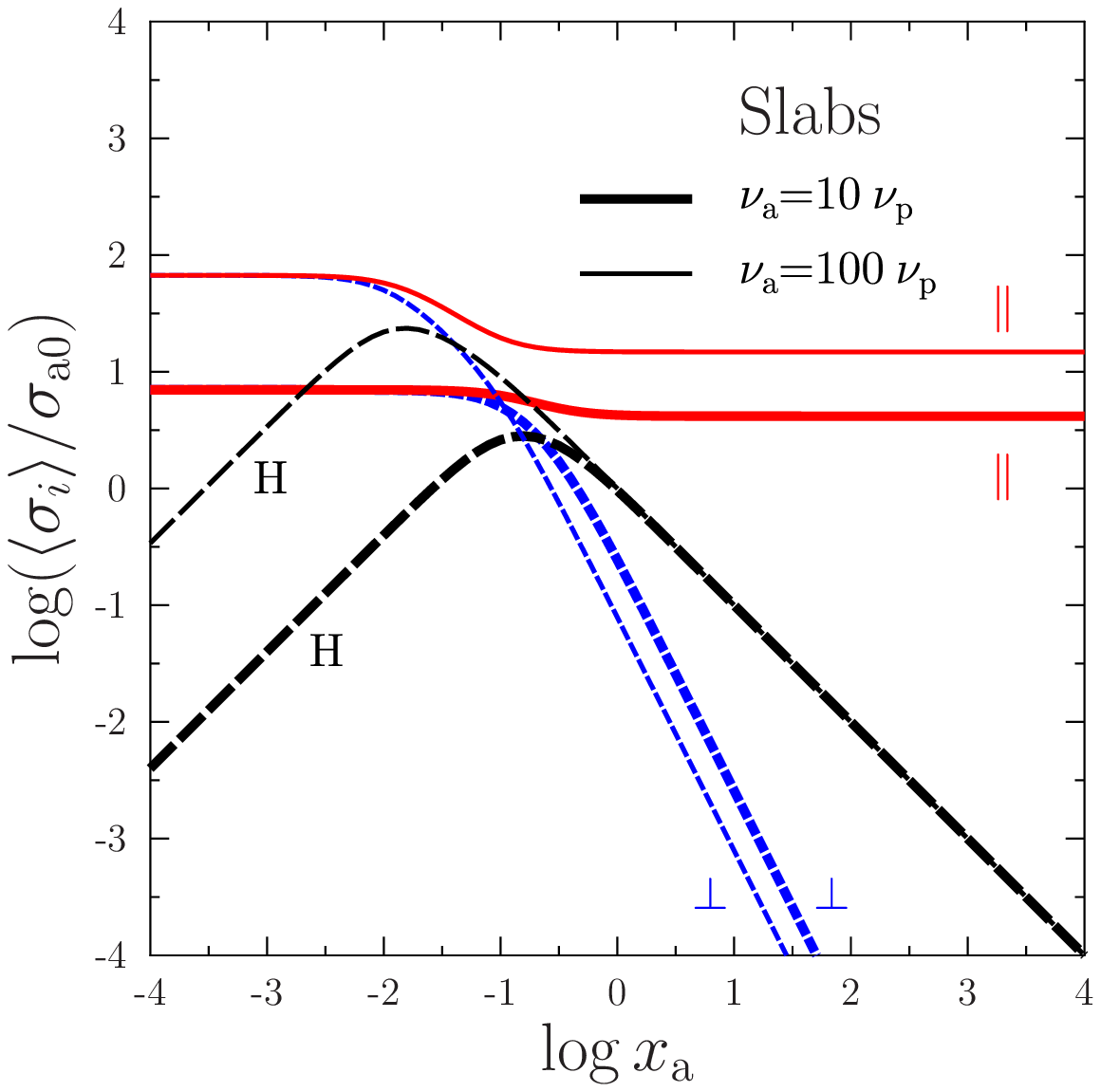}%
\includegraphics[width=0.33\textwidth]{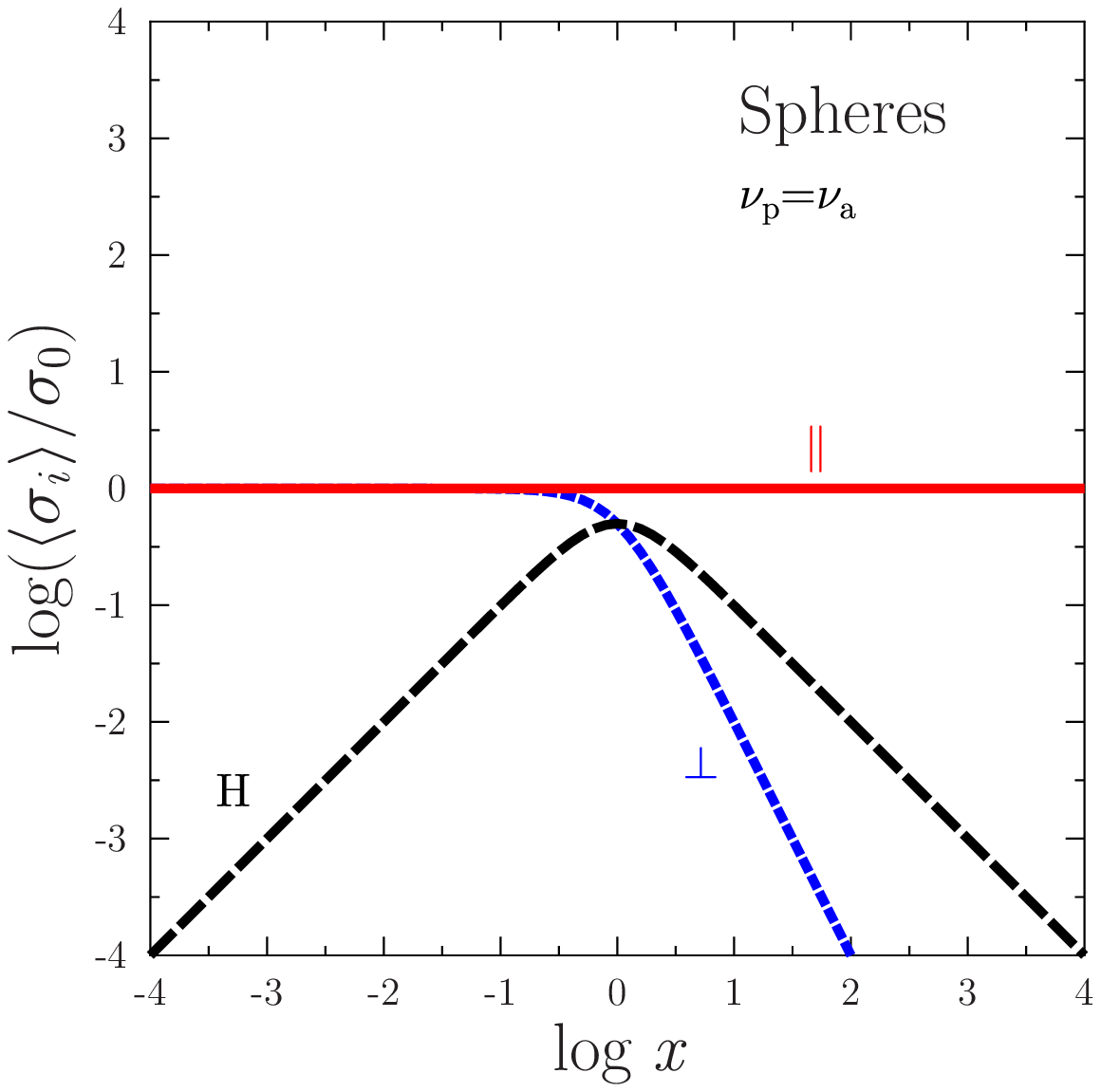}
\caption{(Color online) Electron conductivities $\langle
\sigma_i\rangle$ (averaged over orientations of nuclear
clusters, in units of the leading conductivities
(\ref{e:B=0,sigmaap}) at $B=0$) along $\bm{B}$ ($i=\parallel$, solid
lines), across $\bm{B}$ ($i=\perp$, short-dashed lines), as well as
the Hall conductivity ($i=$H, long-dashed lines) as functions of the
leading Hall parameter for the phases of rods, slabs and spherical bubbles
(from left to right). For rods and slabs the conductivities are
given for two ratios of the leading to weaker collision frequencies,
10 and 100 (thick and thin lines, respectively). See text for
details.} \label{fig:av}
\end{figure*}

\subsection{Illustrative examples}
\label{sec:examples}

By way of illustration, in Fig.\ \ref{fig:rods} we show 
the components of the electron conductivity tensor $\sigma_{\alpha
\beta}$, given by equations (\ref{e:sigma,kappa}) and
(\ref{e:matrixA}), in phases of rods or rod-like bubbles versus the
leading Hall parameter $x_{\rm p}$, equation (\ref{eq:hall-param}).
The components are expressed in terms of the leading (`normal')
conductivity $\sigma_{\rm p0}$ at $B=0$, equation
(\ref{e:B=0,sigmaap}). Recall that the presented ratio
$\sigma_{\alpha \beta}/\sigma_{\rm p0}$ is the same as
$\kappa_{\alpha \beta}/\kappa_{\rm p0}$. The three panels correspond
to the three inclination angles $\theta=0$, 45 and $90^\circ$ of
$\bm{B}$ to the symmetry axis. The ratio of the leading and weaker
electron collision frequencies is taken to be $\nup/\nua=10$. We
present all six different tensor components, $\sigma_{xx}$,
$\sigma_{yy}$ and $\sigma_{zz}$ (short-dashed, solid, and
long-dashed thick lines, respectively), as well as $\sigma_{yx}$,
$\sigma_{zx}$ and $\sigma_{zy}$ (dot-dashed, short dot-dash-spaced
and long dot-dash-spaced thin lines, respectively).

The conductivities plotted in the left-hand ($\theta=0$) and right-hand
($\theta=90^\circ$) panels are described in Section \ref{s:Bneq0}
(equations (\ref{e:B1}) and (\ref{e:B2}), respectively). At $\theta=0$
the  conductivity $\sigma_{zz}$ is unaffected by the field, being
determined by the weaker collision frequency $\nua$. At
$\theta=90^\circ$ the conductivity $\sigma_{xx}$ is not affected by
$\bm{B}$, being determined by $\nup$. If $x_{\rm p} \to 0$, we
reproduce the field-free limit (Section \ref{s:B=0}), with the
leading (`normal') $\sigma_{xx}$ and $\sigma_{yy}$ conductivities 10 times smaller than
the `abnormal' conductivity $\sigma_{zz}$  (because $\nup/\nu_{\rm
a}=10$, in our example).

In the limit of weak electron magnetization the diagonal components
stay nearly constant, while $\sigma_{zx}$ grows as $B^2$ and
all non-vanishing Hall components increase as $B$. In the limit of
high electron magnetization the non-Hall components either stay
constant (as $\sigma_{xx}$, $\sigma_{zz}$ and $\sigma_{zx}$ at $\theta=45^\circ$
in Fig.\ \ref{fig:rods}) or decrease as $B^{-2}$ (like $\sigma_{yy}$
in the same case). In this regime the Hall components
decrease as $B^{-1}$.

Fig.\ \ref{fig:slabs} plots the same conductivities as in
Fig.\ \ref{fig:rods} for $\theta=0$, 45 and 90$^\circ$ but in phase
II of nuclear slabs. The conductivities are expressed in
units of the leading (`normal') conductivity $\sigma_{\rm a0}$ at
$B=0$, equation (\ref{e:B=0,sigmaap}); they are shown as functions
of the leading Hall parameter $x_{\rm a}$. The ratio of the leading
to weaker electron collision frequency is again assumed to be
$\nua/\nup=10$. The main features of the conductivities are
similar to those in Fig.\ \ref{fig:rods} for the
phase of rods. In the regime of weak magnetization for slabs
we also have the abnormal conductivities ($\sigma_{xx}$ and $\sigma_{yy}$) much
higher than the `normal' conductivity ($\sigma_{zz}$).

\section{Averaging over domains}
\label{s:avdomains}

In Section \ref{s:formalism} we have considered the electron
transport in one domain of exotic nuclear structures. This transport
is described by $3 \times 3$ matrices of kinetic coefficients; the
matrix elements behave differently and strongly depend on the
magnetic field $\bm{B}$ and anisotropic properties of nuclear
structures. Macroscopic equations of electron heat and charge
transport contain the conductivity (resistivity) tensors averaged
over an ensemble of domains. In principle, the averaging should take
into account correlation properties of electron transport in
different domains. However, the theory of nuclear pasta phases is
still not very elaborated (Section \ref{sec:intro}), so that no
reliable information has been obtained about predominant
orientations and other statistical properties of the domains.

Therefore, it seems instructive to consider the simplest model of
freely oriented domains. We take the conductivity tensors obtained
in Section \ref{s:formalism} and average them over free orientations
of the domains. It is evident, that after such averaging we obtain
the conductivity tensors for basically isotropic medium in a
magnetic field. In a coordinate frame with the $z$-axis along
$\bm{B}$, the conductivity or resistivity tensors should have the
same form (\ref{e:sph}) as in the phase of spherical bubbles
(Section \ref{s:spherical-bubbles}). Any tensor is characterized by
three components which describe transport along and across $\bm{B}$
as well as the Hall transport, although these components are
generally more complicated than (\ref{e:sph1}). Therefore, it is
sufficient to determine these three components.

To this aim, let us take, for instance, the electric conductivity
tensor (\ref{e:sigma,kappa}) and calculate
$\sigma_\parallel=(\bm{b}\cdot \hat{\sigma} \cdot \bm{b})$,
$\sigma_\perp=(\bm{e}_1\cdot \hat{\sigma} \cdot \bm{e}_1)$ and
$\sigma_{\rm H}=(\bm{e}_1\cdot \hat{\sigma} \cdot \bm{e}_2)$, for a
given orientation of the symmetry axis. Here, $\bm{e}_1$ and
$\bm{e}_2$ are two orthogonal unit vectors perpendicular to
$\bm{b}$, with $\bm{e}_2=\bm{e}_1 \times \bm{b}$. We obtain
\begin{equation}
    \left(
        \begin{array}{c}
     \sigma_\parallel \\
         \sigma_\perp \\
         \sigma_{\rm H}
        \end{array}
        \right)
        =\frac{e^2n_{\rm e}}{m_*\Delta}\,
        \left(
        \begin{array}{c}
        \nua\nup b_x^2+\nup^2 b_z^2+\omega^2 \\
        \nua \nup(e_{1x}^2+e_{1y}^2)+\nup^2 e_{1z}^2 \\
        \nua \omega b_z^2+\nup\omega b_x^2+e_{1z}e_{2z}
         (\nup^2-\nup \nua)
        \end{array}
        \right).
\label{e:sigma123}
\end{equation}
Now the averaging over orientations of nuclear clusters, which we
denote as $\langle \ldots \rangle$, is equivalent to averaging over
rotation of the ($\bm{e}_1$, $\bm{e}_2$) plane about the $\bm{b}$
axis and to subsequent averaging over orientations of the $\bm{b}$
axis. These averagings can be done analytically with the result that
in any reference frame with the $z$ axis parallel to $\bm{B}$
\begin{equation}
   \langle \hat{\sigma}\rangle=
     \left(
     \begin{array}{ccc}
      \langle\sigma_\perp \rangle & -\langle\sigma_{\rm H} \rangle & 0 \\
        \langle \sigma_{\rm H} \rangle & \langle \sigma_\perp \rangle & 0 \\
        0   & 0 & \langle \sigma_\parallel \rangle
     \end{array}
    \right).
\label{e:avsigma}
\end{equation}
The averaged conductivity components are given by
\begin{equation}
    \left(
        \begin{array}{c}
     \langle \sigma_\parallel^{} \rangle \\
         \langle \sigma_\perp^{} \rangle \\
         \langle \sigma_{\rm H}^{} \rangle
        \end{array}
        \right)
        =\frac{e^2n_{\rm e}}{m_*\omega^2}\,
        \left(
        \begin{array}{c}
       (\omega^2+\nup^2) (\omega^2 + \nup \nua ) H - \nup \\
         {1 \over 2} \left( \nua \nup ( \omega^2 - \nup^2) H + \nup \right)  \\
        \omega - \omega \nua \nup^2 H
        \end{array}
        \right) ,
\label{e:sigma123a}
\end{equation}
where
\begin{equation}
   H= \left\{
     \begin{array}{l l}
      (rs)^{-1}\,{\rm arctan}\, (s/r) & {\rm at~~} \nua>\nup,\\
        (2rs)^{-1}\, \ln\, [(r+s)/(r-s)] & {\rm at~~} \nua<\nup,\\
        1/(\nua^3 +\omega^2 \nua) & {\rm at~~} \nua=\nup,
     \end{array}
    \right.
\label{e:H}
\end{equation}
with $r=\sqrt{\nup (\omega^2+\nua \nup )}$ and
$s=\omega \sqrt{|\nup-\nua|}$.

All three conductivities $\langle \sigma_\parallel \rangle$, $\langle
\sigma_\perp \rangle$, and $\langle \sigma_{\rm H} \rangle$ depend
generally on $B$ (or on $\omega$) and on collision frequencies
$\nua$ and $\nup$. In the limit of weak electron magnetization we
naturally have $\langle \sigma_\parallel \rangle =\langle
\sigma_\perp \rangle = \langle \sigma_0 \rangle$, where
\begin{equation}
     \langle \sigma_0 \rangle =
         \frac{e^2 n_{\rm e} \langle \nu^{-1} \rangle}{m_*},
        \quad \langle \nu^{-1} \rangle = \frac{2}{3\nup}+\frac{1}{3\nua};
\label{e:sigma0}
\end{equation}
$\langle \sigma_0 \rangle$ being the field-free averaged conductivity, and
$\langle \nu^{-1} \rangle$ the averaged inverse collision
frequency (averaged relaxation
time) of electrons;  $\langle \sigma_0 \rangle$ is given by the
familiar Drude formula. 
According to
(\ref{e:sigma123}) and (\ref{e:sigma123a}), in the same limit
of weak electron magnetization the Hall term behaves as
\begin{equation}
\langle \sigma_{\rm H} \rangle
\approx \frac{e^2 n_{\rm e}\langle \nu \rangle \, \omega }{m_* \nup^2 \nua }
\propto B,
\label{e:avhall}
\end{equation}
where
$\langle \nu \rangle =(\nua+2\nup)/3$ is the
angle averaged collision frequency.

For illustration, in Fig.\ \ref{fig:av} we show
all three electron conductivities $\langle
\sigma_i \rangle$ averaged over orientations of domains, in units of
the leading (`normal') conductivities at
$B=0$. The conductivities along $\bm{B}$ ($i=\parallel$)
are plotted by solid lines, across $\bm{B}$ ($i=\perp$) by short-dashed lines,
and the Hall conductivities ($i=$H) by long-dashed lines. These
conductivities are presented as
functions of the leading Hall parameters
(\ref{eq:hall-param}) for the phases of rods,
slabs and spherical bubbles (left, middle and right
panels, respectively). For the phases of rods and
slabs the conductivities are presented at two ratios of the leading to
weaker collision frequencies, 10 and 100 (thick and thin lines,
respectively).

As discussed above, in the weak magnetization regime the
conductivities $\langle \sigma_\parallel \rangle$ and
$\langle \sigma_\perp \rangle$ are almost equal and independent of
$\bm{B}$. At large difference between $\nua$ and  $\nup$ these
conductivities are much larger than the corresponding `normal'
conductivities $\sigma_{\rm a0}$ or $\sigma_{\rm p0}$ (Section
\ref{s:formalism}, equation (\ref{e:B=0,sigmaap})). This results
from the presence of abnormally high conductivities in any domain.
By averaging over orientations of domains we inevitably allow
for abnormally high transport in any direction.

In the regime of strong magnetization the longitudinal conductivity, again,
becomes independent of $B$. However, it depend on $\nua/
\nup$ as
\begin{equation}
  \langle \sigma_{\parallel} \rangle
    =\frac{e^2 n_{\rm e}}{m_*}\,S,
\label{e:highBparallel}
\end{equation}
where
\begin{equation}
   S = \left\{
     \begin{array}{l l}
      (\gamma \, \zeta)^{-1}\,{\rm arctan}\,
            (\zeta /\gamma) & {\rm at~~} \nua>\nup, \\
      (2 \gamma \zeta)^{-1}\,
            \ln\, [(\gamma+\zeta)/(\gamma-\zeta)] & {\rm at~~} \nua<\nup , \\
        1/\nua & {\rm at~~} \nua=\nup,
     \end{array}
    \right.
\label{e:S}
\end{equation}
with $\gamma=\sqrt{\nup}$ and $\zeta=\sqrt{|\nup- \nua|}$.
The higher the difference between $\nua$ and $\nup$, the lower $\langle
\sigma_\parallel \rangle$ (measured in units of the leading field-free
conductivity).

In phase of spheres $\langle \sigma_\parallel
\rangle$ is naturally independent of $B$ being equal to $\langle
\sigma_0 \rangle$ at any $B$.

As for the transverse averaged conductivity at high magnetization,
it is suppressed with growing $B$ as $\langle \sigma_\perp \rangle \propto
B^{-2}$ (Fig.\ \ref{fig:av}). The appropriate asymptote is
\begin{equation}
   \langle \sigma_\perp \rangle=
     \frac{e^2 n_{\rm e}\nup }{2 m_*\omega^2}\, (1+\nua S).
\label{e:sigmaperp}
\end{equation}

The averaged Hall conductivity at high magnetization behaves as
$\langle \sigma_{\rm H} \rangle \propto B^{-1}$. It is
given by the universal asymptote
\begin{equation}
 \langle \sigma_{\rm H} \rangle \approx
 \frac{e^2n_{\rm e}}{m_* \omega},
\label{e:sigmaH}
\end{equation}
which is independent of collision frequencies (meaning that $\langle
\sigma_{\rm H} \rangle$ becomes collisionless) and of specific pasta
phase. In this regime, $\langle \sigma_{\rm H} \rangle$ is governed
by classical Hall drift of electrons in crossed electric and
magnetic fields (e.g., \citealt{KINETICS81}). Generally, the
anisotropy of collision frequencies has larger effect on averaged
conductivity components in phase of slabs than in phase of rods.

Finally, let us discuss the tensor of averaged resistivity
$\langle \hat{\cal R} \rangle$ which has the same form as
(\ref{e:sph}). A simple averaging of the resistivity tensor
(\ref{e:hatR}) gives
\begin{equation}
    \langle {\cal R}_\parallel \rangle
=\langle {\cal R}_\perp \rangle=\langle {\cal R}_0 \rangle
=\frac{ \langle \nu \rangle m_*}{e^2 n_{\rm e}},\quad
\langle {\cal R}_{\rm H} \rangle
=\frac{ \omega m_*}{e^2 n_{\rm e}}.
\label{e:resist}
\end{equation}
This tensor has the same structure as the
tensor (\ref{e:sph1}) in phase of spheres. The only collision
frequency involved here is the averaged collision frequency $\langle
\nu \rangle$. It determines the averaged longitudinal and transverse
resistivities which are equal and independent of $\bm{B}$. The Hall
resistivity $\langle {\cal R}_{\rm H} \rangle$  
exceeds $ \langle {\cal R}_0 \rangle$ at high magnetization,
$\omega >\langle \nu \rangle$.

Let us stress, that, generally, $\langle \hat{\cal R} \rangle
\neq \langle \hat{\sigma} \rangle^{-1}$ and the difference can be
substantial in anisotropic media. While calculating Joule heat
(\ref{e:joule}) one should definitely use the averaged resistivity,
but not the inverse averaged conductivity. Large difference between
$\nua$ and $\nup$ does not produce any significant difference
between $\langle {\cal R}_0 \rangle$ and
the leading non-averaged resistance ($1/\sigma_{\rm a0}$ or $1/\sigma_{\rm p0}$).

\section{Collision frequencies}
\label{s:coll-freq}

So far, we have treated the collision frequencies $\nua$ and $\nup$ as given.
Let us discuss how to calculate them.

\subsection{Spherical bubbles}
\label{s:holes}

\begin{figure}
\includegraphics[width=0.45\textwidth, bb= 0 0 348 348]{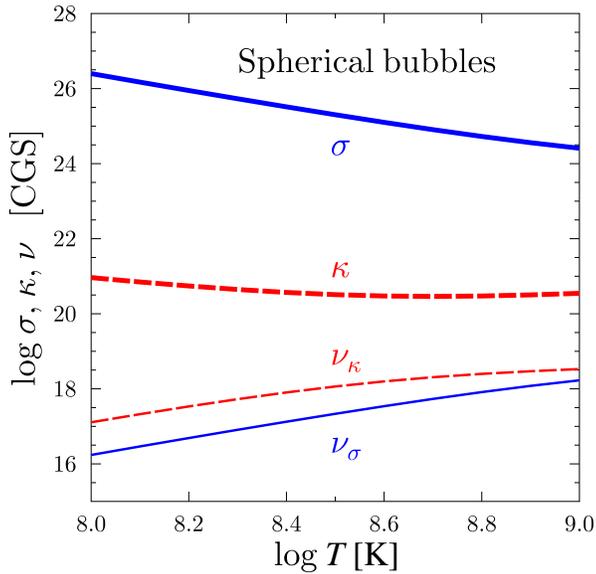}
\caption{(Color online) Electric and thermal conductivities (thick
solid and dashed lines, respectively) versus temperature in phase IV of spherical
bubbles at $\rho=1.423 \times 10^{14}$ g~cm$^{-3}$.
Thin solid and dashed lines show the effective
collision frequencies $\nu_\sigma$ and $\nu_\kappa$ for electric and
thermal conductivities, respectively. See text for details.
}
\label{fig:holes}
\end{figure}

The formalism for calculating the collision frequency $\nu_0$
($=\nua=\nup$) in phase IV of spherical bubbles is the same as that
for spherical nuclei in the ordinary neutron star crust. For
illustration, consider model I of \citet{Oyamatsu93}, according to
which spherical bubbles exist in the narrow density range (from
about $1.42\times 10^{14}$ to $1.43\times 10^{14}$ g~cm$^{-3}$,
Section \ref{s:general}), just near the interface between the mantle
and neutron star core. Analytic fits for proton and neutron
density profiles in nuclear bubbles are given by \citet{HPY07} in
their Appendix B.

For certainty, we take $\rho=1.423 \times 10^{14}$ g~cm$^{-3}$
(i.e., the baryon number density $n_{\rm b}=0.08517$ fm$^{-3}$)
although the results are almost independent of density within such a
narrow density range. The radius of a spherical Wigner-Seitz cell
around one spherical bubble is $R_{\rm WS}=15.4$ fm. The cell
contains $A_{\rm tot}\approx 1300$ nucleons including $Z_{\rm
tot}\approx 45$ protons. They are mainly free neutrons
distributed almost uniformly. The neutron and proton density
profiles within the cell decrease near the cell's center. This
creates neutron and proton holes which can be collectively treated
as a nuclear hole. The minimum neutron density in the center is
only slightly lower than the average neutron density, while the
central proton density drops to zero. The proton hole is equivalent
to vacancies of $Z_{\rm h}\approx 11$ protons; whereas the entire
nuclear hole is equivalent to vacancies of $A_{\rm nuc}\approx 48$
nucleons. The root-mean-squared radius of the proton hole is about
8.2 fm; the neutron hole is shallow but larger. These nuclear holes
behave as charged quasiparticles and form a strongly coupled Coulomb
system. The electrons are distributed almost uniformly and scatter
off charge fluctuations of proton vacancies. The electron chemical
potential is $\mu \approx 87.3$ MeV. The electron transport
properties can be calculated by applying standard formalism of
electron transport in a plasma of spherical nuclei
\citep{POT99,GYP01,POT13} and replacing the nuclei by the nuclear
holes.

\begin{figure}
\includegraphics[width=0.45\textwidth]{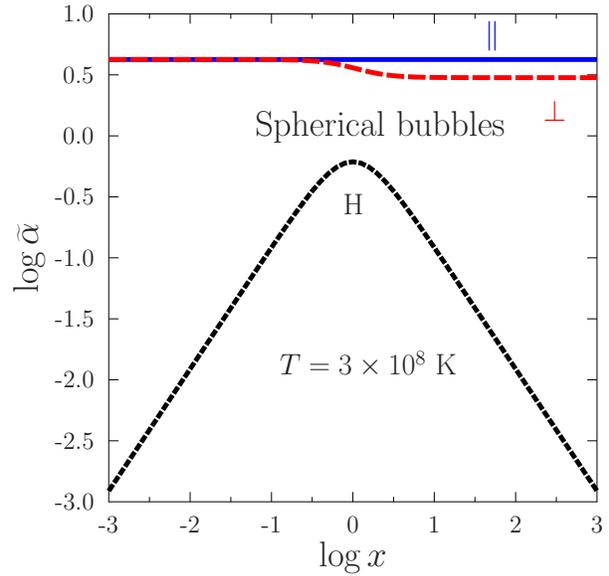}
\caption{(Color online) Reduced dimensionless thermopower components 
 $\widetilde{\alpha}_\parallel$,
$\widetilde{\alpha}_\perp$ and $\widetilde{\alpha}_{\rm H}$ (solid,
long-dashed and short-dashed lines, respectively) given by
(\ref{e:thermosa}) versus the Hall parameter $x=\omega/\nu_\sigma$
in phase IV of spherical bubbles at $\rho=1.423
\times 10^{14}$ g~cm$^{-3}$ and $T=3 \times 10^8$ K. See text for details. }
\label{fig:thermoholes}
\end{figure}

Fig.\ \ref{fig:holes} shows the temperature dependence of the electric and
thermal conductivities, $\sigma$ and $\kappa$ (thick lines),
in the phase of spherical bubbles
at $\rho=1.423 \times 10^{14}$ g~cm$^{-3}$. The
conductivities are calculated with the method of
quasi-potential for electron-nucleus scattering suggested by
A. Y. Potekhin \citep{POT99,GYP01} and modified to consider electron
scattering by nuclear holes. The calculations
include Coulomb coupling of nuclear holes,
Debye--Waller factor, multi-phonon processes, and shape of spherical
bubbles. The results are more advanced than those based
on the simplified model (\ref{e:tau-nu}) of the collision integral in the
relaxation time approximation. The displayed conductivities are
solely produced by electron-hole scattering; other scattering
mechanisms (e.g., the electron-electron scattering,
\citealt{SY06}) are disregarded. Low-temperature suppression
of the electron-phonon scattering due to electron band structure
effects \citep{CHUG12} is not included.

The conductivities $\sigma$ and $\kappa$ in Fig.\ \ref{fig:holes}
are expressed in CGS units (in s$^{-1}$ and
erg~cm$^{-1}$~s$^{-1}$~K$^{-1}$, respectively). They are remarkably
close to those calculated for standard models of inner neutron star
crust of spherical nuclei (without nuclear pasta, e.g.
\citealt{GYP01,POT13}; note that the conductivities $\sigma$ in
fig.\ 7 of the latter paper are plotted -- from top to bottom -- for
$\log T$=8, 8.5 and 9, respectively). Two thin lines in Fig.\
\ref{fig:holes} present the effective collision frequencies
$\nu_\sigma$ and $\nu_\kappa$ (expressed in s$^{-1}$) which describe
the electric and thermal conduction, respectively (see, e.g.,
Potekhin et al. 1999). At $T \gtrsim 10^9$~K these frequencies
almost coincide. This means that the collision integral in the
Boltzmann transport equation can be accurately described in the
relaxation time approximation, equation (\ref{e:tau-nu}). At lower
temperatures $\nu_\kappa$ becomes higher than $\nu_\sigma$ and the
relaxation time approximation is less accurate. However, at
temperatures $T \gtrsim 10^8$~K, most important for many
applications, this approximation can still be used, as least for
semi-quantitative analysis.

Having the effective collision frequency, we can easily estimate
magnetic fields $B=B_{\rm H}$ which correspond to the Hall parameter
$x= \omega/\nu=1$. They are characteristic magnetic fields which
affect the electron transport. For $T=10^8$ and $10^9$~K with
$\nu=\nu_\sigma$ we obtain $B_{\rm H}\sim 2 \times 10^{11}$ and $2
\times 10^{13}$~K, respectively. The $B$ fields expected in neutron
star interiors can be much higher. The lower the temperature, the
lower the collision frequency, and the lower the fields which affect
the electron transport.

So far we have not discussed the thermopower tensor $\hat{\alpha}$.
For nuclear bubbles, we can present this tensor in the same form as
$\hat{\sigma}$, equation (\ref{e:sph}), and write
$\hat{\alpha}=[\pi^2 k_{\rm B}^2 T/(3ep_{\rm F}v_{\rm
F})]\,\hat{\widetilde{\alpha}}$, where $\hat{\widetilde{\alpha}}$ is
the reduced (dimensionless) thermopower and $v_{\rm F}\approx c$ is
the electron Fermi velocity. From equation (\ref{e:thermopower}) one
obtains \citep{UY80}
\begin{equation}
  \widetilde{\alpha}_\parallel=3+{\cal P},\quad
  \widetilde{\alpha}_\perp=3+\frac{{\cal P}}{1+x^2},\quad
  \widetilde{\alpha}_{\rm H}=\frac{x \,{\cal P}}{1+x^2},\quad
\label{e:thermosa}
\end{equation}
where ${\cal P}={\partial \ln\, (\epsilon
\nu_\sigma(\epsilon))^{-1}}/{\partial \ln p}\,|_{p=p_{\rm F}}$ and,
again, $x=\omega/\nu_\sigma=B/B_{\rm H}$. Note that the expression
for $\widetilde{\alpha}_\parallel$ (which is independent of $B$) is
valid as long as the electrons are strongly degenerate, regardless
of the applicability of the relaxation time approximation
(\ref{e:tau-nu}). In contrast, the expressions for
$\widetilde{\alpha}_\perp$ and $\widetilde{\alpha}_{\rm H}$ depend
on $B$ through $x$ and require the approximation
(\ref{e:tau-nu}).

Fig.\ \ref{fig:thermoholes} presents all thermopower components for
the phase of nuclear bubbles at $\rho=1.423 \times 10^{14}$
g~cm$^{-3}$ and $T=3 \times 10^8$~K as a function of $x=B/B_{\rm
H}$, with $B_{\rm H}\approx 2.03 \times 10^{12}$~G. The thermopower
$\widetilde{\alpha}_\parallel$ along the magnetic field is shown by
the solid line, while $\widetilde{\alpha}_\perp$ and
$\widetilde{\alpha}_{\rm H}$ are plotted by the long-dashed and
short-dashed lines, respectively. Recall that the thermopower
components, contrary to the components of $\hat{\sigma}$,
$\hat{\kappa}$ and $\hat{\cal R}$, can change sign depending on
electron scattering mechanism. Signs of the thermopower components
are regulated by the parameter ${\cal P}$. As shown by \citet{UY80},
neglecting the effects of the Debye-Waller factor and finite sizes
of atomic nuclei, one has ${\cal P}=-1$ for scattering of
ultrarelativistic degenerate electrons in high-temperature crystals
(when the relaxation time approximation is strictly valid). Indeed, we
reproduce ${\cal P}=-1$ at $T=10^9$~K when the
relaxation time approximation is really accurate (see above).
However, taking into account the Debye-Waller factor and finite
sizes of the nuclei, we obtain ${\cal P}\approx 1.1$ (instead of
--1). Therefore, at the bottom of the neutron star crust the effects
of the Debye-Waller factor and of the shape of proton (proton
hole) charge distribution are very important. If we decrease the
temperature from $10^9$ to $10^8$ K, the relaxation time
approximation becomes less accurate and ${\cal P}$ increases slowly from
1.1 to 1.3.

According to Fig.\ \ref{fig:thermoholes}, the magnetic field
modifies $\widetilde{\alpha}_\perp$ but not very strongly. However,
it significantly affects $\widetilde{\alpha}_{\rm H}$ which
regulates the Hall drift of the magnetic field. At $x \lesssim 1$
the Hall thermopower directly determines the drift velocity
\citep{UY80}. Actually, the thermopower can be large enough to
affect generation and evolution of magnetic fields in hot neutron
stars.

\subsection{Non-spherical nuclear structures}
\label{s:non-spheric}

Calculations of $\nua$ and $\nup$ for phases I--III of non-spherical
nuclear structures are much more complicated. In principle, they can
be performed using the technique similar to that for spherical nuclei.
To this aim one needs detailed studies of proton charge fluctuations
and their effects on electron scattering. For instance, 
useful information on vibration properties of nuclear clusters can be 
obtained using their elastic constants (e.g., \citealt{PETPAL1998}).
Proton charge fluctuations for the phase of slabs have been analyzed 
briefly by \citet{WIS00}. Classical molecular dynamics
simulations of dynamical behavior of nuclear pasta have been
conducted by \citet{HOROWITZ2005}. However, these results seem insufficient
for a reliable determination of $\nua$ and $\nup$.
The required physical ingredients are the effective
structure factor $S(\bm{q})$, Debye-Waller factor and nuclear
form-factor $F(\bm{q})$. Because nuclear clusters are
non-spherical, these quantities will strongly depend on orientation
of wave vector $\bm{q}$, leading to essential difference
between $\nua$ and $\nup$. To the best of our knowledge, the 
collision frequencies $\nua$ and
$\nup$ for non-spherical clusters have not been calculated.

In this situation, it seems reasonable to assume that the leading
effective frequency, $\nua$ or $\nup$, is close to that for models
of neutron star matter with spherical nuclei, while the other
(abnormal) effective frequency is smaller. The ratio of the leading
to the smaller frequency can be considered as a free 
parameter, ranging within reasonable limits,  for instance, from
10 to 100. It would be difficult to be more exact using existing
calculations. For illustration, one may look at fig 3 of \citet{SY06}
which shows partial thermal conductivities of electrons at $B=0$ 
in a neutron star crust of spherical nuclei. Assume the presence
of nuclear pasta at the crust bottom and take the leading collision frequency
the same as for electron--nucleus scattering in fig 3. Let the weaker
collision frequency be limited by electron-electron collisions (whose
contribution is also plotted in fig 3). Then 
in the temperature range $10^8-10^9$~K the ratio of the leading to weaker
collision frequencies would be $\sim 10$. 
Accurate microscopic calculations of $\nua$ and $\nup$
would be highly desirable.

\section{Conclusions}
\label{s:conclude}

We have analyzed electron transport in possible nuclear
pasta phases in a mantle of magnetized neutron star. Specifically,
we have performed model calculations of the tensors of electric and
thermal conductivity, $\hat{\sigma}$ and $\hat{\kappa}$,
thermopower, $\hat{\alpha}$, and electric resistivity $\hat{\cal
R}$. The model assumes the presence of domains containing
(quasi)-ordered exotic nuclear structures of one type (rods, slabs,
rod-like bubbles, or spherical bubbles; Table \ref{tab:pasta}). The
collision integral in the electron transport equation is described
in the generalized relaxation time approximation
(\ref{e:collision_integral}) with two, generally different,
effective collision frequencies, $\nua$ and $\nup$, along and across
the symmetry axis of corresponding nuclear clusters. For 
non-spherical nuclear clusters one can expect the
presence of the leading (normal) collision frequency ($\nua$ or
$\nup$) and abnormally weak frequency (Table \ref{tab:pasta}),
whereas  for spherical bubbles $\nua=\nup$. Our model
is simple but it allows us to include the effects
of strong magnetic fields which would be difficult to do using
more complicated models.

In Section \ref{s:formalism} we have studied the electron transport
in one domain and have found it rather complicated, with six
generally different components of conductivity tensors.

In Section \ref{s:avdomains} we have considered electron transport
in matter containing freely oriented domains of
one type. The tensors of kinetic coefficients 
(\ref{e:avsigma}), (\ref{e:sigma123a}), and (\ref{e:resist}) averaged over
orientations of the domains are calculated analytically and have
much simpler structure than their non-averaged counterparts. Each
tensor contains only three different components, along and across
$\bm{B}$, as well as the Hall component. These components show
sufficiently simple dependence on $\bm{B}$. Nevertheless, this
dependence is less trivial than in ordinary basically isotropic
media with $\nua=\nup$. For instance the angle-averaged longitudinal
conductivities $\langle \sigma_\parallel \rangle$ and $\langle
\kappa_\parallel \rangle$ depend now on $B$ and can be much higher
than the leading normal conductivities (\ref{e:B=0,sigmaap}) for $B=0$ 
determined by the
leading collision frequencies. This enhancement of averaged
longitudinal conduction is produced by the contribution of
abnormally high conduction with lower collision frequency. The 
enhancement contradicts some expectations (e.g., \citealt{PONS2013,
DISORDER15}) that nuclear pasta suppresses the conduction. It seems both
possibilities (suppression and enhancement) can be realized; see below.  

In Section \ref{s:coll-freq} we have outlined how to calculate the
collision frequencies $\nua$ and $\nup$. The
collision frequencies and transport properties in phase of nuclear
bubbles can be calculated rather reliably; they appear similar to
those in traditional phase of spherical nuclei. However, the
collision frequencies for non-spherical nuclear clusters
are uncertain and require further attention.

Much work should be done to clarify transport properties of
nuclear pasta. The major problem is to reliably calculate effective
collision frequencies $\nua$ and $\nup$ for non-spherical nuclear
clusters I--III (as outlined in Section \ref{s:non-spheric}). The
problem requires accurate studies of proton charge fluctuations in
these structures.

We have treated nuclear clusters in one domain as ordered and
having the same shapes. If the ordering is not exact and the
parameters (e.g., $\nua$ and $\nup$) strongly
vary, an additional averaging over disorder and parameter
distributions has to be performed. We have assumed that orientations
of domains are free. If not, more complicated averaging over
orientations is needed.

We have neglected quantum oscillations of electron kinetic coefficients in
very strong magnetic fields due to population of new electron Landau levels
with increasing density (e.g., \citealt{PALEX1999}). 
If the electrons occupy many Landau levels, these
oscillations are not too strong and our results should reproduce
oscillation-averaged kinetic properties. In addition, since nuclear
clusters are loosely bound, very strong magnetic fields may
influence proton charge fluctuations and affect transport properties
in this way. All these problems are beyond the scope of this
paper.

Another, more serious problem would be to account for possible
impurities and defects in nuclear pasta. It seems that these
effects can greatly enhance or suppress electron transport.

Finally, let us stress, that there are other heat and charge
transport mechanisms in nuclear pasta which should be taken into
account along with the mechanisms studied above. For instance, one
should bear in mind thermal conductivity due to electron-electron
collisions (e.g. \citealt{SY06}) and due to superfluid phonons 
\citep{SFPHONONS2009} associated
with the presence of free superfluid neutrons.

The results of the present studies can be useful for modeling of
thermal relaxation between crust and core in young neutron stars and
also in accreting neutron stars in X-ray transients with the crust
overheated during accretion episodes (e.g., \citealt{PONS2013}). 
In addition, the results can
be used to study evolution of magnetic fields
in neutron stars and related phenomena (e.g., \citealt{DISORDER15}).

\section*{Acknowledgments}
This work was supported by the Russian Science Foundation,
grant 14-12-00316.

\end{document}